# A comparison of three heart rate detection algorithms over ballistocardiogram signals

Ibrahim Sadek[1,2*], Bessam Abdulrazak[1]

[1]AMI-Lab, Computer Science Department, Faculty of Science, University of Sherbrooke, Sherbrooke, QC, Canada

[2]Biomedical Engineering Dept, Faculty of Engineering, Helwan University, Helwan, Cairo, Egypt

## Abstract

Heart rate (HR) detection from ballistocardiogram (BCG) signals is challenging because the signal morphology can vary between and within-subjects. Also, it differs from one sensor to another. Hence, it is essential to evaluate HR detection algorithms across several datasets and under different experimental setups. In this paper, we studied the potential of three HR detection algorithms across four independent BCG datasets. The three algorithms were as follows: the multiresolution analysis of the maximal overlap discrete wavelet transform (MODWT-MRA), continuous wavelet transform (CWT), and template matching (TM). The four datasets were obtained using a microbend fiber optic sensor, a fiber Bragg grating sensor, electromechanical films, and load cells, respectively. The datasets were gathered from: a) 10 patients during a polysomnography study, b) 50 subjects in a sitting position, c) 10 subjects in a sleeping position, and d) 40 subjects in a sleeping position. Overall, CWT with derivative of Gaussian provided superior results compared with the MODWT-MRA, CWT (frequency B-spline), and CWT (Shannon). That said, a BCG template was constructed from DataSet1. Then, it was used for HR detection in the other datasets. The TM method achieved satisfactory results for DataSet2 and DataSet3, but it did not detect the HR of two subjects in DataSet4. The proposed methods were implemented on a Raspberry Pi. As a result, the average time required to analyze a 30-second BCG signal was less than one second for all methods. Yet, the MODWT-MRA had the highest performance with an average time of 0.04 seconds.

## 1. Introduction

Remote monitoring of vital signs, i.e., body temperature, heart rate (HR) [1], blood oxygen saturation [2], respiratory rate (RR), and blood pressure, has attracted public health attention due to rapidly emerging infectious diseases, e.g., coronavirus disease [3]. Besides, changes in vital signs are critical in assessing the severity and prognosis of epidemic diseases. Specifically, these changes are significant signs of deteriorating patient health and thus present an opportunity for early detection and intervention. In hospital practice, nursing staff, and doctors rely on intermittent vital signs, usually measured every 8-hour shift. Hence, early deterioration indicators could be overlooked, particularly at night, when deterioration could progress undetected until the next morning [4]. Bed-embedded ballistocardiogram (BCG) sensors have presented encouraging results for detecting particular vital signs, namely HR and RR [5,6].

---

[*] Corresponding author.
E-mail address: ibrahim.sadek@usherbrooke.ca (I. Sadek)

Additionally, these sensors have been implemented for diagnosing severe sleep disorders, specifically sleep apnea [7–9]. BCG-based sensors can be used for remote monitoring of vital signs without disturbing users' daily activities [10]. In contrast, wearable sensors such as actigraphs [11] can impose limits on users, especially for vulnerable populations with behavioral symptoms. For example, if the sensor is not waterproof, it has to be taken off before showering. Moreover, if the sensor has a short battery life, it needs to be taken off frequently for charging. These situations will inevitably cause inconvenience to patients and likewise disrupt the data collection [12]. The choice between wearable and non-wearable sensors should be made to cope with each patient group's medical conditions. There will always be a compromise between the continuity of data and patient convenience. Although BCG sensors can help alleviate some shortcomings of wearable sensors, they are highly prone to motion artifacts, e.g., body movements.

Furthermore, they can only be practical for observing patients in a single bed setting. That is to say; these sensors are not designed to deliver spot readings for substituting nurse observations. However, they are intended for monitoring trends in vital signs, taking into account their capacity to collect longitudinal data [4]. Various signal processing and machine learning algorithms have been suggested to scrutinize BCG signals (Figure 1), considering the multiple factors that affect the signal quality. The goal of these algorithms is to automatically identify the "J" peak of the "I-J-K" complex [5]. BCG is "*a technique that records the movements of the body imparted to it by the forces linked to the contraction of the heart and acceleration and deceleration of blood as it is ejected and moved in the large vessels*" [13]. Under controlled conditions, if the subject sleeps on the bed without movement, this peak can be detected using a classical peak detector.

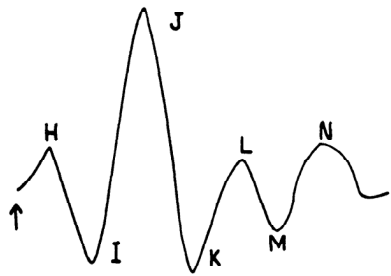

Figure 1 Diagram of a typical BCG signal with letters used to identify the component parts. The arrow indicates the beginning of electrical ventricular systole [14].

Nonetheless, this is not conceivable in real-life scenarios. The sensor location is another element that can largely affect the signal quality. Ideally, the closer the sensor is to the chest and abdomen region, the better the signal quality. Hence, the sensor's desired location is under the upper part of the body, in which it can be placed under the bed sheet or the mattress. In real-life scenarios, we cannot predict subjects' sleep positions, and thus, unless the bed is covered entirely by pressure sensors, the signal quality can be highly degraded. Still, this arrangement will increase the deployment's cost. Furthermore, the BCG signal morphology can vary from one sensor to another and between and within patients. These restrictions should always be considered when a system is designed for analyzing BCG signals [15,16]. Fast Fourier transform (FFT), Hilbert transform, template matching, autocorrelation, cepstrum, wavelet transform, and empirical mode decomposition, among others, have been implemented for automatic HR detection from BCG signals [5,17,18].

Moreover, convolutional neural networks (CNNs) have been employed to segment the "I-J-K" complexes and detect HR in BCG signals [19–21]. Although CNNs yielded satisfactory results, the training was performed in a controlled setting with a small sample size, including only healthy individuals. Besides, BCG signals were recorded over a short time, i.e., 35 seconds [20]. Wavelet analysis, in particular, has proved to be a valuable tool in analyzing BCG signals because of its ability to handle nonlinear and nonstationary physiological signals [17,18].

To our knowledge, most of the proposed approaches were not evaluated across different datasets or under different experimental setups. In other words, the generalization of one method across different datasets or settings is not yet feasible. This problem occurs because BCG signals are not benchmarked, as is the case with electrocardiogram signals. As a result, proposed methods are mainly applied to proprietary data.

To this end, this research aims to compare three HR detection algorithms across four independent BCG datasets acquired under different experimental setups. These algorithms include the Maximal Overlap Discrete Wavelet Transform (MODWT), Continuous Wavelet Transform (CWT), and Template Matching (TM). The objective of this comparative study is to examine the efficacy of each method across the various datasets and under different experimental setups. Also, we evaluate whether a BCG template from one dataset can be used to detect HR in the other datasets. The rest of the paper is structured as follows. Section 2 discusses related approaches that use wavelet transform or template matching for HR detection from BCG signals. Section 3 describes the experimental setup and data collection protocols. Also, it outlines the principles of the three proposed methods. Findings and contributions made are presented in Sections 4 and 5. Finally, the paper is concluded in Section 6.

## 2. Related Work

The wavelet transform (WT) aims at decomposing the signal into smooth and detail components. Thus, the component (or the sum of multiple components), including the most similar pulse-like peaks, can be adopted to locate the J-peaks of the BCG signal.

## 2.1. Wavelet Based Methods

Table 1 presents a summary of wavelet-based methods used in the literature to detect HR from BCG signals. Zhu et al (2005, 2006) [22,23] applied the "à trous" WT to raw sensor data acquired via a liquid pressure sensor under a pillow. The signals were gathered from 13 healthy subjects during sleep for about two hours. Motion artifacts caused by head and body movements were reduced by a threshold method using the raw signals' standard deviation (SD). The "Cohen-Daubechies-Feauveau" 9/7 (CDF 9/7) biorthogonal wavelet was selected for designing the decomposition and reconstruction filters. The $4^{th}$ and $5^{th}$ detail components were realigned in the signal phases, and their amplitudes were summed to estimate the BCG signal. Finally, J-peaks were detected using a modified Pan-Tompkins algorithm [24] after noise reduction with a soft threshold method.

Jin et al (2009) [25] employed a translation-invariant WT based on adaptive threshold wavelet shrinkage for signal denoising. The Symlet wavelet of order 8 (sym8) was adopted for detecting cardiac cycles because it was closer in shape to the BCG signal. The signal was collected from a healthy subject, but there was no information on the data acquisition process. Then, J-peaks were detected using a pseudo-period detection approach that can locate the signal's largest swings [26]. Postolache et al (2010) [27] designed a framework for measuring HR from two EMFi sensors embedded in the seat and backrest of a wheelchair. BCG signals were gathered from eight subjects seated in the chair over 15 minutes. At first, signals were denoised using discrete stationary WT combined with a soft threshold method. Secondly, the denoised signals were decomposed via a discrete WT—Daubechies (db5) wavelet function. In addition, the cardiac signal was reconstructed by summing the $8^{th}$, $9^{th}$, and $10^{th}$ detail components. At last, a time-domain peak detection algorithm was used to detect J-peaks. A similar approach was introduced by Pino et al (2015) [28], in which BCG signals were acquired via two EMFi sensors embedded in the seat and backrest of a typical chair. Raw sensor data were collected from 19 subjects in a laboratory for over 1 minute and 35 subjects in a hospital waiting area for over 2 minutes. Daubechies (db6) wavelet function was used for the decomposition, and the cardiac signal was reconstructed by summing the $4^{th}$ to $7^{th}$ detail components. J-peaks were detected using a customized peak detector algorithm.

Gilaberte et al (2010) [29] proposed to use CWT to detect HR from subjects standing on a bathroom scale. Six subjects participated in the study, and data were recorded over ten seconds in different days and conditions (i.e., before and after meals). Subjects were instructed not to talk or move to eliminate noise. The cardiac signal was located using Daubechies (db10) wavelet function at different scale ranges. The authors suggested that two ranges must be explored in the case of very different HR values. Alvarado-Serrano et al (2016) [30] implemented CWT with B-splines to detect HR using data from subjects in a sitting position. A piezo-electric sensor was fixed to a typical chair seat, and raw sensor data were gathered from seven subjects for about 100 seconds. The $5^{th}$ scale of CWT was defined as the optimal scale for HR detection. J-peaks were detected through learning and decision stages. In these stages, several experimental parameters were determined that could limit their use in another dataset.

Table 1 Summary of wavelet-based approaches used to detect HR from BCG signals. "CDF": Cohen-Daubechies-Feauveau; "sym": Symlet; "db": Daubechies; "D": detail-component; "s": seconds; "min": minutes.

| Authors | Subjects | Sensor | Environment | Acquisition time | Wavelet | Wavelet function | Cardiac signal |
|---|---|---|---|---|---|---|---|
| **Zhu et al (2005, 2006)** [22,23] | 13 | Liquid pressure sensor | Sleep lab | 120 min | DWT | CDF 9/7 | Sum(D4:D5) |
| **Jin et al (2009)** [25] | 1 | N/A | Lab | N/A | DWT | sym8 | N/A |
| **Postolache et al (2010)** [27] | 8 | EMFi sensor | Lab | 15 min | DWT | db5 | Sum(D8:D10) |
| **Pino et al (2015)** [28] | 35 | EMFi sensor | Lab and hospital | 2 min | DWT | db6 | Sum(D4:D7) |
| **Gilaberte et al (2010)** [29] | 6 | Strain gauges | Lab | 10 s | CWT | db10 | N/A |
| **Alvarado-Serrano et al (2016)** [30] | 7 | Piezo-electric sensor | lab | 100 s | CWT | B-splines | Scale 5 |

## 2.2. Template Matching Based Methods

Shin et al (2008) [31] proposed to use a TM approach for BCG beat detection. BCG signals were recorded using three sensors: air-mattress, loadcell, and EMFi-film. An individual BCG template was constructed for each sensor using ensemble averaging of valid BCG cycles centered at J-peak points. Five records of 30-second were gathered for each sensor, and the matching was performed using the correlation coefficient function. Paalasmaa et al (2015) [32] presented a method for detecting interbeat intervals from BCG signals acquired with a piezo-electric sensor. A BCG template was created using a clustering-based method. Then, interbeat intervals were detected via a customized cross correlation-based approach. The BCG template was continually being updated based on the detected interbeat intervals. Raw sensor data were recorded overnight from 40 patients in a sleep clinic and 20 subjects at their homes. Nevertheless, only 46 overnight recordings were used in the study. Cathelain et al (2019) [33] introduced a similar approach to [31]. However, the matching was achieved using dynamic time wrapping. In this study, a Murata SCA11H BCG sensor was deployed, and data were acquired from ten healthy subjects over 20 to 50 minutes long naps. The initial BCG template was updated with the newly detected J-peaks to alleviate the variabilities in the BCG signal shape.

## 3. Methodology

### 3.1. Experimental Setup and Data Collection

Figure 2 illustrates the experimental setup of the four datasets. The first dataset (DataSet1) was acquired using a microbend fiber optic sensor (MFOS) from 10 sleep apnea patients. The patients underwent polysomnography (PSG) at the sleep laboratory of Khoo Teck Puat Hospital, Singapore (elapsed time: $8.12 \pm 0.54$ hours). The MFOS was placed under the patient's mattress in the upper part of the bed. The PSG electrocardiogram (ECG) signals were used as a gold-standard to assess the proposed HR detection methods. For more details about the MFOS and dataset, readers are referred to [7].

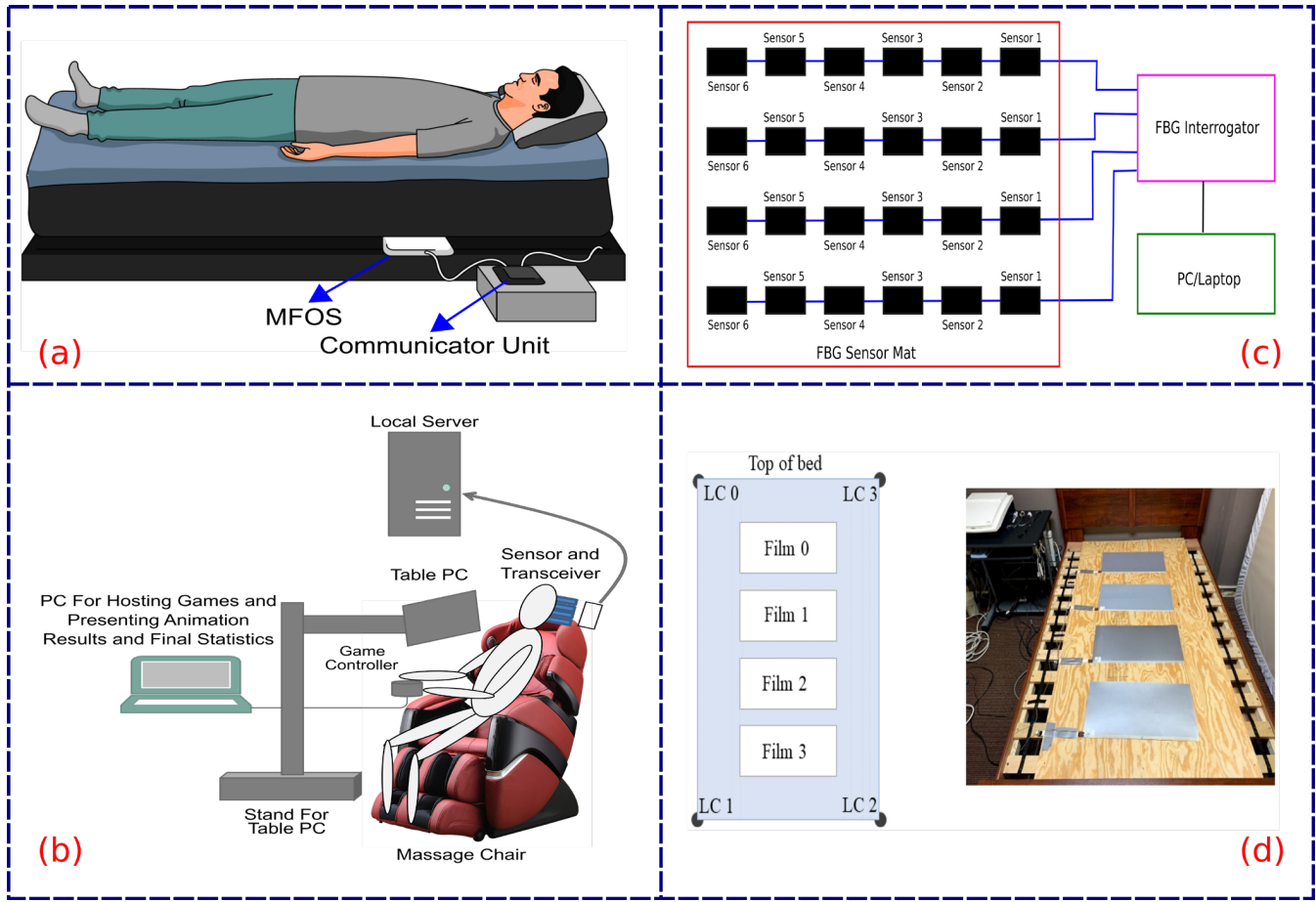

Figure 2 Diagram describing the experimental setup of the four datasets. The top left figure shows the location of the MFOS under a typical bed mattress, the bottom left figure shows the location of the MFOS on the headrest of a massage chair, the top right shows the structure of the FBG sensor mat, and the bottom right shows the location of the EMFis and load cells [34].

The second dataset (DataSet2) was collected in a realistic setting by an MFOS from 50 participants sitting in a massage chair (elapsed time: $1.01 \pm 0.11$ hours). The MFOS was installed on the chair's headrest, and BCG signals were transmitted wirelessly to a computer via Bluetooth. The study aimed to evaluate the participants' stress levels at various time points. The participants underwent a sequence of stress-induced activities, rest (no-activity), and relaxation therapy [35,36]. The continuity of contact was a significant issue in this study. If the participants had lifted or relocated their heads, we could not have recorded the BCG signals. We manually discarded participants' data with artifacts severe enough to degrade BCG signal quality in light of this issue. Therefore, we could only analyze data from 39 participants. ECG signals were simultaneously recorded, and they were used as a reference for HR detection.

The third dataset (DataSet3) was gathered from 10 subjects using a sensor mat combining four fiber Bragg grating (FBG) sensor arrays. Each sensor array consisted of six sensors, and they were set to collect signals at 250Hz. The sensor arrays were placed under the subjects' a) head, b) chest, c) chest and abdomen, and d) under hip. The experiment was split into two phases: 10 minutes of supine sleeping and 10 minutes of side sleeping. The ideal sensor's location was under the chest and abdomen. Similarly, optimal results were obtained by averaging signals from the six sensors in the time domain [37]. As a result, the fused signal was employed for HR detection.

The fourth dataset (DataSet4) was recently published by Carlson et al (2020) [34]. The signals were obtained from 40 subjects using: a) four electromechanical films (EMFis) and b) four load cells. The EMFis were placed underneath the mattress, and the load cells were positioned under the bedposts. Overall, 4.5 hours of data were collected. The EMFis and load cells were set to collect data at a sampling rate of 1 kHz. For the EMFis, HR was computed by fusing the four EMFis signals using a pairwise maximum operation. Average fusion was also examined. Nevertheless, the fused signal was distorted, and the main features of a typical BCG signal (i.e., "I-J-K" complexes) were missed. For the load cells, HR was detected by fusing signals from two load cells denoted as "LC2" and "LC3" using a pairwise maximum operation.

Across the four datasets, acquired BCG signals were preprocessed to separate motion artifacts and no-activity intervals. The preprocessing step was carried out using a sliding time window of 30-seconds with an overlap of 15-seconds. The standard deviation (SD) of each time window was computed. Then, the median absolute deviation (MAD) of the SDs was calculated. Time windows with SD greater than 2 times the MAD were considered as motion artifacts. Furthermore, time windows with SD less than a fixed value (10 mV) were regarded as no-activity intervals and discarded from further analysis. No-activity implies that no pressure force was applied to the MFOS. The remaining time windows were considered informative signals wherein BCG signals could be extracted. Following the preprocessing step, a bandpass filter was applied to the informative signals (artifact-free) for obtaining BCG signals. The bandpass filter was constructed by cascading high and low pass filters as follows: 1) second-order Chebyshev type I high-pass filter with a maximum ripple of 0.5 dB, and a critical frequency of 2.5Hz followed by 2) fourth-order Chebyshev type I low-pass filter with a maximum ripple of 0.5dB and critical frequency of 5Hz.

### 3.2. Maximal Overlap Discrete Wavelet Transform

Unlike DWT, the MODWT skips the downsampling after filtering the signal. The reason is that it gains other features, e.g., invariant to time-shifting, the ability to analyze any time series with arbitrary sample size, and increased resolution at a coarser scale. Besides, it generates a more asymptotically efficient wavelet variance estimator than DWT [38,39]. MODWT decomposes a signal into a number of details and a single smooth. The details describe variations at a particular time scale, whereas the smooth describes the low-frequency variations.

Given a time series $X_t$ of $N$ samples, the level $J_0$ MODWT is a transform consisting of $J_0 + 1$ vector that is, $\widetilde{W}_1, \cdots, \widetilde{W}_{J_0}$ and $\widetilde{V}_{J_0}$. All these vectors have a dimension $N$. The vector $\widetilde{W}_j$ comprises wavelet coefficients linked to changes on the scale $\tau_j = 2^{j-1}$, whereas the $\widetilde{V}_{J_0}$ comprises the MODWT scaling coefficients linked to averages on the scale $\lambda_{J_0} = 2^{J_0}$ [40]. The $\widetilde{W}_j$ and $\widetilde{V}_j$ can be constructed by filtering $X_t$ as follows:

$$\widetilde{W}_{j,t} = \sum_{l=0}^{N-1} \tilde{h}_{j,l}^\circ X_{t-l \bmod N}, \tag{1}$$

$$\widetilde{V}_{j,t} = \sum_{l=0}^{N-1} \tilde{g}_{j,l}^\circ X_{t-l \bmod N}, \tag{2}$$

$t = 0, \cdots, N-1$ and $j = 1, 2, \cdots, L$, where $\tilde{h}_{j,l}^\circ$ and $\tilde{g}_{j,l}^\circ$ are the $j$th level MODWT wavelet and scaling filters (high- and low-pass filters) obtained by periodizing $\tilde{h}_{j,l}$ and $\tilde{g}_{j,l}$ to length $N$. These filters can be defined by renormalizing the DWT wavelet and scaling filters such as $\tilde{h}_{j,l} = h_{j,l}/2^{j/2}$ and $\tilde{g}_{j,l} = g_{j,l}/2^{j/2}$. The multiresolution analysis (MRA) of the MODWT breaks up a signal into high-pass filtered detail components and a low-pass filtered smooth component. The MRA of the MODWT can be expressed as follows:

$$X = \sum_{j=1}^{L} D_j + S_j, \tag{3}$$

$$D_{j,t} = \sum_{l=0}^{N-1} \tilde{h}_{j,l}^\circ \, \widetilde{W}_{j,t+l \bmod N}, \tag{4}$$

$$S_{j,t} = \sum_{l=0}^{N-1} \tilde{g}_{j,l}^\circ \, \widetilde{V}_{j,t+l \bmod N}, \tag{5}$$

Where $D_j$ is the wavelet detail at decomposition $j$, and $S_j$ is the wavelet smooth at decomposition $j$. Figure 3 shows an example of the MODWT multiresolution analysis for a 10-second BCG signal. It can be seen that the 4[th] level smooth coefficient (S4) represents the J-Peaks of the BCG signal. We briefly discuss the CWT in the next subsection.

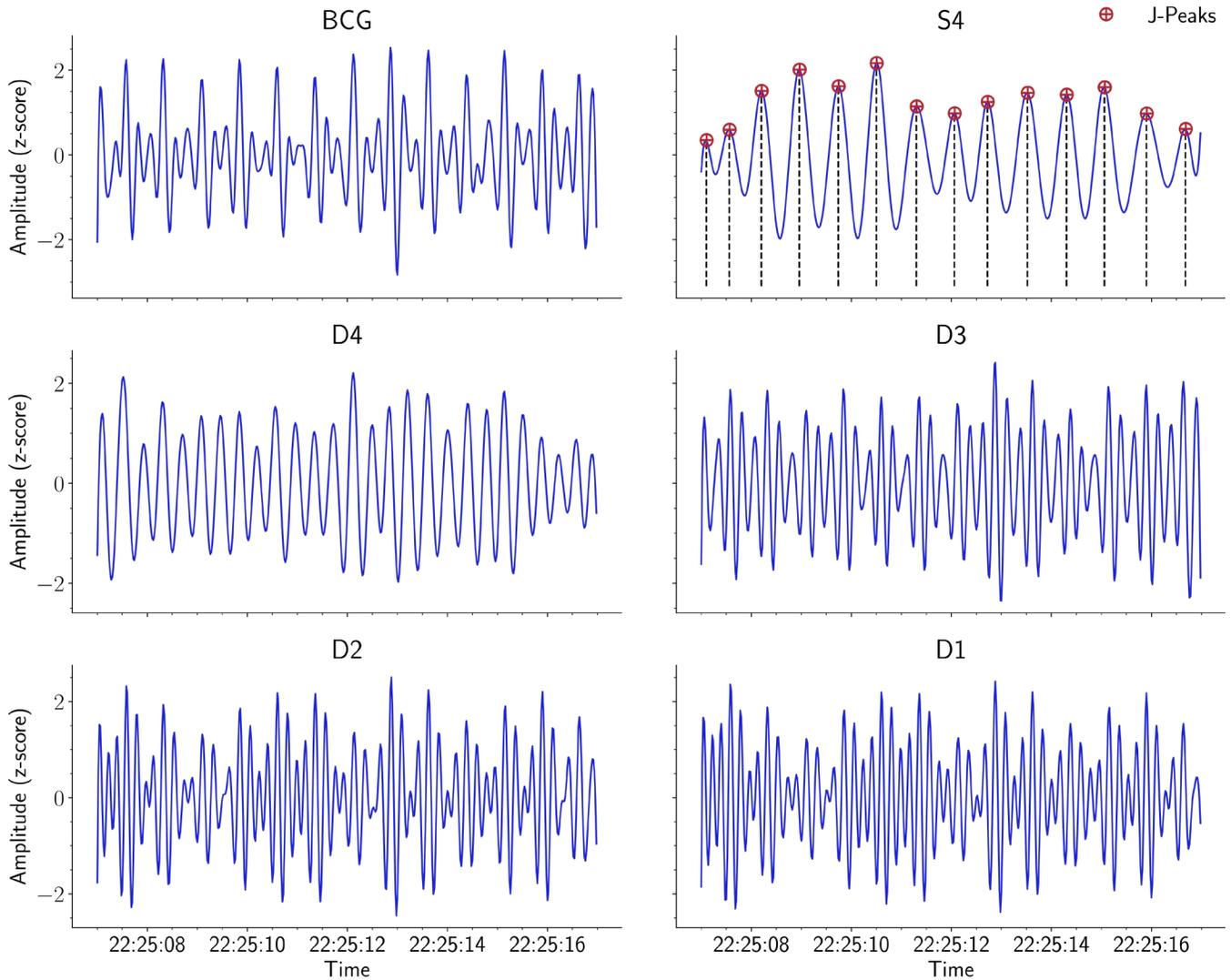

Figure 3 MODWT multiresolution analysis for a 10-second BCG signal. Wavelet Biorthogonal 3.9 (bior3.9) with 4 decomposition levels were opted to analyze the BCG signal. The maximum peaks of the 4[th] level smooth coefficient (S4) correspond to the J-Peaks. The amplitude was normalized (z-score) for better visualization.

### 3.3. Continuous Wavelet Transform

Continuous wavelet transform (CWT) is a time-frequency (more correctly, a time-scale) transform that is a useful tool for examining nonstationary signals. CWT is a generalization of the short-time Fourier transform (STFT) commonly used to analyze nonstationary signals at multiple scales [41]. In a similar way to STFT, CWT applies an analysis window, i.e., a wavelet, to extract segments from a signal. In contrast to STFT, the wavelet is not only translated but dilated and contracted to consider the scale of the activity under consideration. The wavelet's dilation and contraction serve two purposes, i.e., increasing the CWT's sensitivity to long- and short-time scale events, respectively. Given a continuous input signal $x(t)$, the CWT can be defined as follows:

$$C(a,\tau) = \int \frac{1}{\sqrt{2}} \psi\left(\frac{t-\tau}{a}\right) x(t) dt, \tag{6}$$

Where $\psi(t)$ is the mother wavelet, $a$ is a scale, $\tau$ is a shift parameter; $C(a,\tau)$ is a bivariate function obtained by mapping $x(t)$ and a wavelet scaled by $a$ at a given time $\tau$. The localized correlation in time is determined over an integral starting with $t = \tau$ and ending duration $t = \tau + L$, where $L$ is the wavelet's duration. It is noteworthy that short-term events (high-frequency signal components) such as spikes and transients can be determined when the wavelet is contracted ($a < 1$), whereas long-time events (low-frequency signal components) such as baseline oscillations can be determined when the wavelet is stretched ($a > 1$) [41,42]. The result of the CWT can be shown in a graph known as a scalogram. It can be created by estimating the correlation between a signal and wavelets with different scales and then plotting how the correlation of each wavelet changes over a given period [41]. Figure 4 shows a CWT example for a 10-second BCG signal. Gaus2 wavelet was opted to analyze the signal, and the wavelet coefficients at scale 20 (scales 1 to 30) were used to detect the J-Peaks.

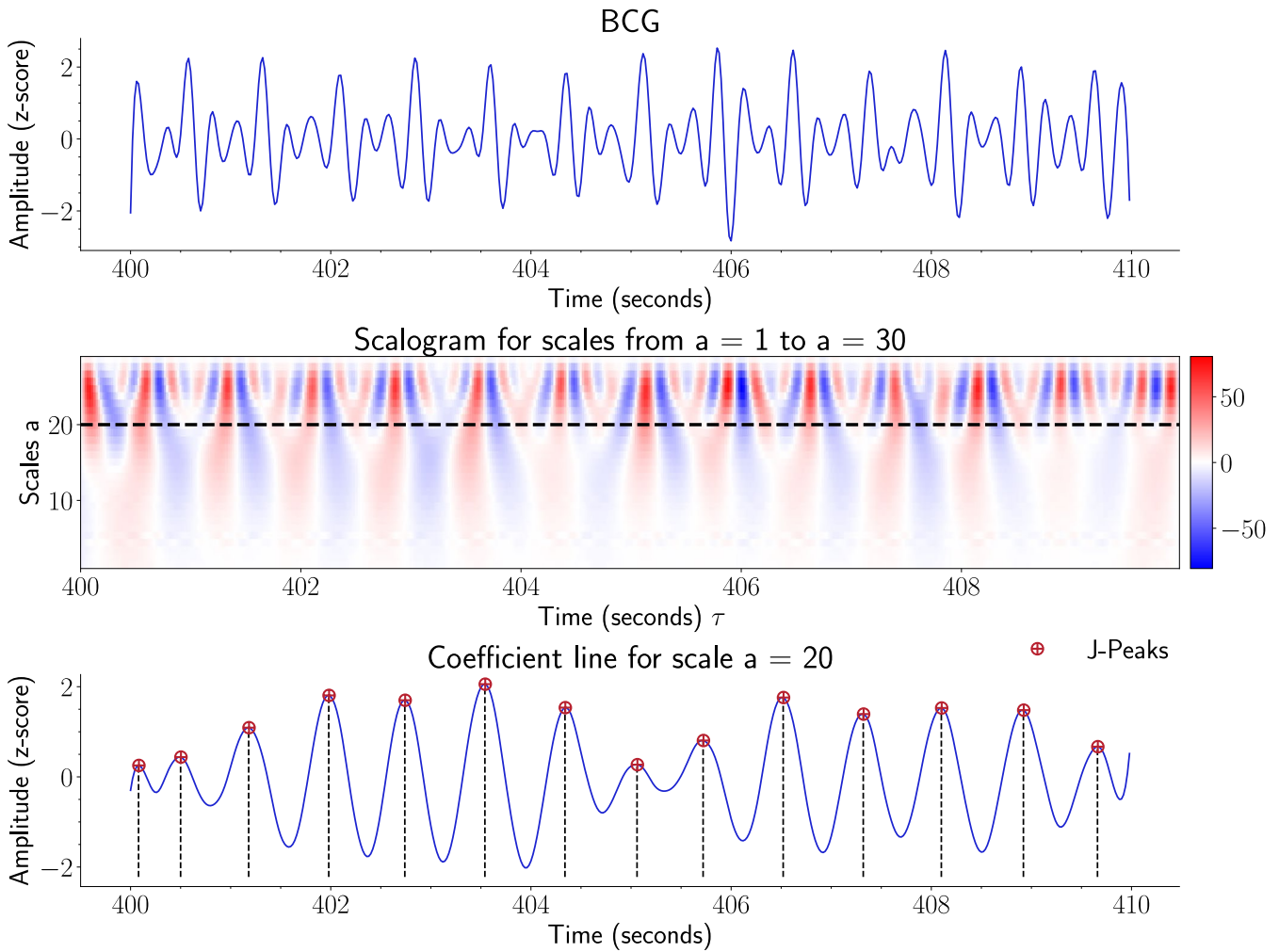

Figure 4 An example of a BCG signal (top), scalogram (center), and coefficients line at scale 20 (bottom). Gaus2 wavelet was designated for analyzing the BCG signal. The dashed black line on the scalogram was the scale (i.e., scale 20) where the J-Peaks were detected. The amplitude was normalized (z-score) for better visualization.

### 3.4. Template Matching

Over the past few decades, template matching (TM) has been one of the most common methods in pattern recognition [43]. This method aims to determine the similarity between a template/prototype signal and a target signal. The main challenge of this method is to choose the prototype and the similarity measure. The prototype (cardiac cycle) was constructed from DataSet1 due to the close contact between the MFOS mat and the participants. However, in DataSet2, BCG signal morphology was primarily affected by the frequent movement of the massage chair. We specified the prototype as follows. **Firstly**, a human expert selected high-quality BCG segments with a size of 30-seconds (1500-samples) using the BCG signals obtained from the 10 patients. This process was performed using a semi-automatic approach, i.e., motion artifacts were automatically detected using the preprocessing step described in Section 3.1. Then, each 30-second BCG segment was manually verified. The term "high-quality" implies that the segment does not include any signs of motion artifacts. Also, cardiac cycles can be easily identified. **Secondly**, we divided each 30-second segment into equal slices of 1-second (50-samples) with an overlap of 0.5-second. The redundancy created by the overlapped slices enabled us to accurately detect cardiac cycles, considering the relatively small sampling frequency of the MFOS. **Thirdly**, we manually classified each 1-second slice into BCG signal and non-BCG signal based on the presence of the "I-J-K" complex using a custom app with a GUI in MATLAB. The non-BCG slices were discarded from our analysis. **Finally**, the prototype was constructed by an ensemble averaging the valid slices centered at J-peak points (Figure 5 and Figure 6).

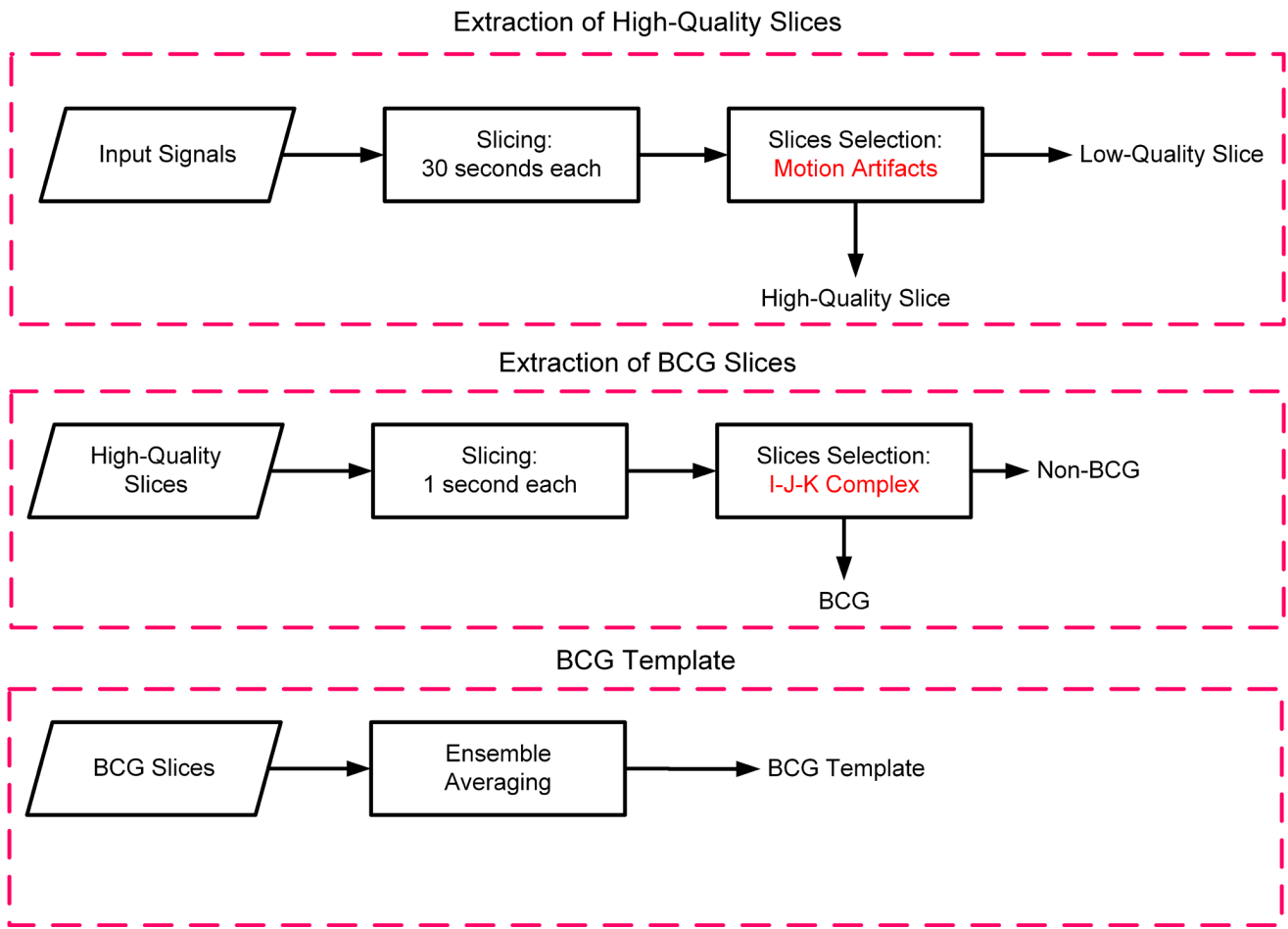

Figure 5 Block diagram describing how the BCG prototype was created.

To this end, the minimum peak distance (MPD) used to detect HR was considered 0.3 seconds, and it was selected using experimental observation. To illustrate, several peak distances were evaluated, i.e., ranging from 0.2 to 0.7 seconds with a step size of 0.05 seconds (Figure 7). The MPD was appointed by examining the effects of two measures, i.e., precision (Prec) and mean absolute error (MAE), on HR detection across the 10 patients. Detected HR values were classified into correct and incorrect detections for each MPD. Then, Prec was calculated to provide "*a rough estimate of how a large portion of the detected HR values are correct,*" that is, "*how correct the detected*

*HR values are*" [32,44,45]. It was calculated as follows: $Prec = correct/(correct + incorrect)$. The average MAE (i.e., between true and correctly predicted HR values) in beats per minute (BPM) tended to increase with increasing the distance. In addition, the average precision tended to decrease with increasing the distance. Therefore, the 0.3-second interval was assigned as an optimal interval to strike a balance between lower MAE (5.02) and higher precision (68.91%).

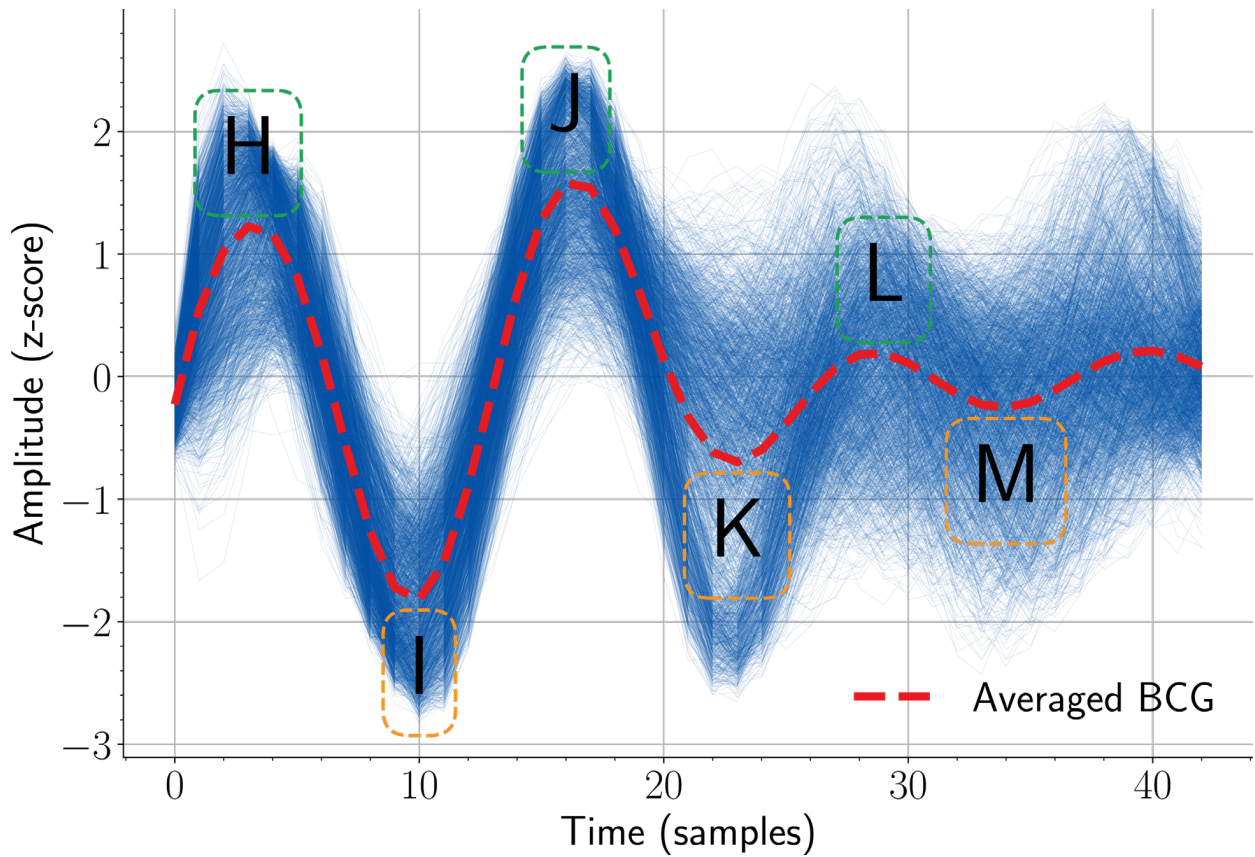

Figure 6 An ensemble averaging of BCG signals. The "I-J-K" represents the ejection phase of the cardiac cycle.

For each cardiac cycle, a candidate J-peak was detected by finding the maximum peak of the cross-correlation function (CCF) between the template and the BCG signal.

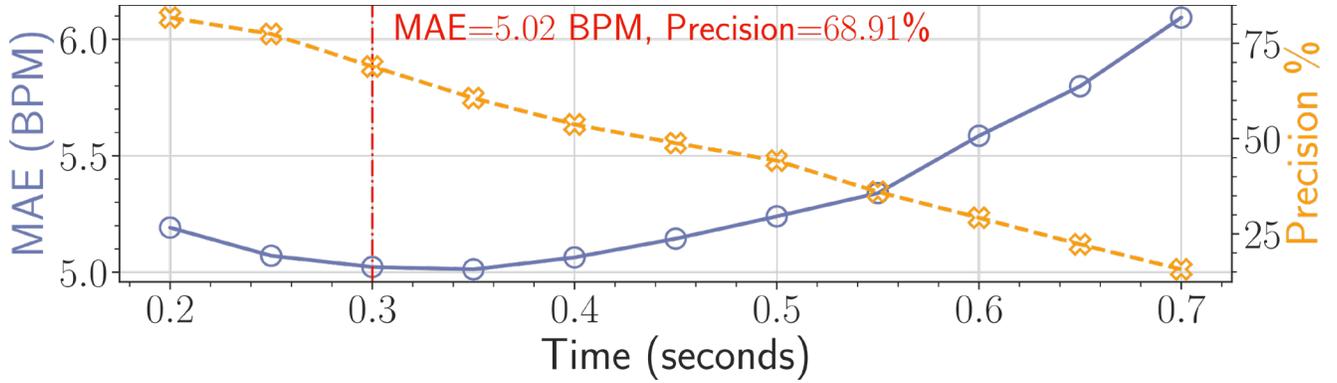

Figure 7 HR detection performance metrics (i.e., MAE and Prec) against the minimum peak distances.

CCF is defined by calculating the correlation coefficients between the samples of the template ($x$) and the BCG signal shifted by $k$, $(y(n - k))$ [46]. The formula is as follows:

$$\rho_{xy}(k) = \frac{\frac{1}{N}\sum_{n=0}^{N-1}(x(n) - \bar{x}) \cdot (y(n - k) - \bar{y})}{\sqrt{\left(\frac{1}{N}\sum_{n=0}^{N-1}(x(n) - \bar{x})^2\right) \cdot \left(\frac{1}{N}\sum_{n=0}^{N-1}(y(n) - \bar{y})^2\right)}}, \quad (7)$$

Both signals were supposed to have $N$ samples each (50 samples). At last, J-peaks were only deemed to be heartbeats, if the minimum distances between peaks were 0.3 seconds.

### 3.5. Heart Rate Detection

The HR was computed on a 30-second window and sliding the window by 15 seconds. The time window choice was based on previous studies [7,36,47], taking into account the sampling rate of the sensor (50Hz). Regarding the MODWT-MRA, the Biorthogonal 3.9 (bior3.9) wavelet was appointed to detect cardiac cycles. The bior3.9 wavelet proved to be the most suitable to characterize the profile of cardiac cycles across different wavelets, precisely Daubechies 1 (db1), Symlet 2 (sym2), Coiflets 1 (coif1), and Reverse Biorthogonal 3.1 (rbior3.1) [18]. BCG signals were analyzed using 4 decomposition levels, and the 4th level smooth coefficient was employed for J-peaks detection [7,36,47]. The periodicity of the smooth coefficient reflected the same periodicity as the HR (Figure 3). At last, J-peaks were localized using a peak detector.

Table 2 Parameters and scales of the CWT for HR detection; $P$ is an order-dependent normalization constant, $M$ is the spline order, $B$ is the bandwidth, and $C$ is the center frequency.

| Wavelet | Parameters | Scales |
| --- | --- | --- |
| GausP | $P = 2$ | Range(1, 30), HR scale: 20 |
| FbspM-B-C | $M = 2, B = 1, C = 1$ | Range(1, 100), HR scale: 45 |
| ShanB-C | $B = 1.5, C = 1$ | Range(1, 100), HR scale: 75 |

For CWT, Gaussian Derivative (GausP), Frequency B-Spline (FbspM-B-C), and Shannon (ShanB-C) wavelets were tested for HR detection. $P$ is an order-dependent normalization constant, $M$ is the spline order, $B$ is the bandwidth, and $C$ is the center frequency. For each wavelet, BCG signals were analyzed at different scales using the scalogram (Figure 4), and then the scale reflecting the same periodicity as the HR was designated for J-peaks detection [29]. The designated parameters and scales of the three wavelets are given in Table 2. For the rest of the paper, Gaus2, Fbsp2-1-1, and Shan1.5-1.0 will be used to refer to the CWT wavelets. On the other hand, BCG signals obtained from DataSet1 were used to construct a BCG template (training phase). The created template was then employed to detect HR in the remaining datasets, as outlined in subsection 3.4. The HR value at a time $t_n$, at which the $nth$ maximum occurred, was defined as follows:

$$HR_n = \frac{60}{t_n - t_{n-1}}, \qquad (8)$$

Where $t_n$ is the time at $nth$ local maxima and $t_{n-1}$ is the time at $(n-1)th$ local maxima in the designated MODWT coefficient or CWT scale.

The proposed methods were implemented using Python 3.8 on a Raspberry Pi 3 Model B (Quad-Core 1.2GHz and 1GB RAM) as an embedded system. The MODWT-MRA method was applied using the "*wmtsa-python*" library[1], while CWT-based methods were applied using the "*Scaleogram*" library[2]. In short, the average time needed to analyze a 30-second BCG signal was less than one second for the 5 methods. Yet, the MODWT-MRA took less time compared to other methods, i.e., 0.04 seconds. Shan1.5-1.0 and Fbsp2-1-1 required more time to analyze a BCG signal, i.e., 0.43 and 0.44 seconds (Table 3). This performance was expected because a large number of scales were used, specifically 100. The time required for the TM was 0.12 seconds. The improved performance for the MODWT-MRA occurred because the "*wmtsa-python*" library is written in Python and Cython. However, "*Scaleogram*" library uses an adaptive convolution algorithm selection; that is, the scale processing loop switches to FFT-based convolution when the complexity is better in $N * log2(N)$. MODWT-MRA is expected to be more efficient for applications that require real-time processing of the data considering its improved performance. Further information about acquiring raw sensor data from the MFOS can be found in the Appendix.

Table 3 The average time (in seconds) for analyzing a 30-second BCG signal using MODWT-MRA, Gaus2, Fbsp-2-1-1, Shan1.5-1.0, and TM methods.

| Method | Time (seconds) |
|---|---|
| MODWT-MRA | 0.04 |
| Gaus2 | 0.06 |
| Fbsp2-1-1 | 0.43 |
| Shan1.5-1.0 | 0.44 |
| TM | 0.12 |

---

[1] https://gitlab.com/Cimatoribus/wmtsa-python

[2] https://github.com/alsauve/scaleogram

# 4. Results

This section presents the results of the three proposed methods across the four datasets. For each method, the BPM error between the reference ECG and the measurement device was evaluated separately using the MAE, mean absolute percentage error (MAPE), and root mean square error (RMSE). All figures were generated using Python (*Matplotlib, Plotnine, Bokeh, and Seaborn*). The BPM error is reported such as "mean (SD)" unless otherwise stated. Table 9 shows all the errors reported for each of the datasets and each of the algorithms used.

## 4.1. Performance Evaluation of Heart Rate Detection: DataSet1

As presented in Table 4, MODWT-MRA achieved overall MAE, MAPE, and RMSE of 4.71 (1.07), 7.61% (1.65%), and 5.59 (1.02), respectively. The smallest and largest error values were 3.13, 4.55%, 4.05 and 7.16, 9.99%, 7.83 for patients 1 and 8, respectively. Also, both patients had the highest and lowest precision, i.e., 96.53% and 30.77%. The past medical history of patient 8 indicated hypertension and dyslipidemia. Furthermore, this patient had severe obstructive sleep apnea (OSA) with an apnea-hypopnea index (AHI) of 78.2 [7].

Table 4 HR detection performance metrics for DataSet1 using MODWT-MRA.

| Metrics | Patients Ids | | | | | | | | | | Mean (SD) |
|---|---|---|---|---|---|---|---|---|---|---|---|
| | 1 | 2 | 3 | 4 | 5 | 6 | 7 | 8 | 9 | 10 | |
| MAE | 3.13 | 4.92 | 5.06 | 4.70 | 3.78 | 4.68 | 5.00 | 7.16 | 3.87 | 4.83 | 4.71 (1.07) |
| MAPE (%) | 4.55 | 7.45 | 8.76 | 8.15 | 5.90 | 8.66 | 9.07 | 9.99 | 6.31 | 7.26 | 7.61 (1.65) |
| RMSE | 4.05 | 5.81 | 5.94 | 5.60 | 4.68 | 5.66 | 5.88 | 7.83 | 4.69 | 5.74 | 5.59 (1.02) |
| Prec (%) | 96.53 | 81.87 | 81.27 | 83.66 | 94.11 | 70.54 | 85.99 | 30.77 | 94.19 | 83.23 | 80.22 (19.01) |

The BCG signal's quality was poorly affected by these health problems. Thus, reported Prec was low compared with other patients. Furthermore, this low Prec value triggered a large variability in the overall Prec, with a mean and standard deviation (SD) of 80.22% and 19.01%. Regarding CWT-based methods, Gaus2 and Shan-1.5-1.0 provided comparable results to MODWT-MRA (Table 5). Nevertheless, Gaus2 performed slightly better than MODWT-MRA in which the overall MAE, MAPE, and RMSE were 4.71 (1.22), 7.58% (2.17%), and 5.58 (1.20). More importantly, Prec value for patient 8 was highly improved from 30.77% to 45.98%. On the other hand, Fbsp2-1-1 attained the smallest and largest values of the error measures for patients 1 and 7, i.e., 2.27, 3.32%, 3.07 and 6.76, 11.48%, 7.51, respectively. The Prec value for patient 7 was 31.75% that was comparable to the Prec of patient 8 when the MODWT-MRA was applied. It is worth mentioning that patient 7 had been diagnosed with hypertension and severe OSA with an AHI such as 76.6 [7]. Compared to Gaus2 and Shan-1.5-1.0 wavelets, smaller Prec values of 47.22%, 45.73%, and 39.67% were obtained for patients 3, 4, and 6 despite their moderate OSA (i.e., AHI were 23, 27, and 33, respectively). These findings suggested that Fbsp2-1-1 could be more susceptible to patients' comorbidities. As a result, cardiac cycles were not appropriately captured for various time intervals, triggering a total Prec of 69.57% (25.91%).

Table 5 HR detection performance metrics for DataSet1 using CWT-Gaus2, CWT-Fbsp2-1-1, and CWT-Shan1.5-1.0.

| Method | Metrics | Patients Ids | | | | | | | | | | Mean (SD) |
|---|---|---|---|---|---|---|---|---|---|---|---|---|
| | | 1 | 2 | 3 | 4 | 5 | 6 | 7 | 8 | 9 | 10 | |
| Gaus2 | MAE | 2.67 | 4.33 | 5.58 | 5.42 | 3.35 | 4.81 | 5.92 | 6.70 | 4.18 | 4.10 | 4.71 (1.22) |
| | MAPE (%) | 3.88 | 6.51 | 9.52 | 9.30 | 5.25 | 8.61 | 10.61 | 9.17 | 6.86 | 6.11 | 7.58 (2.17) |
| | RMSE | 3.51 | 5.25 | 6.44 | 6.34 | 4.22 | 5.84 | 6.74 | 7.40 | 5.03 | 5.04 | 5.58 (1.20) |
| | Prec (%) | 98.35 | 89.75 | 73.11 | 72.63 | 96.45 | 58.45 | 68.43 | 45.98 | 93.57 | 91.62 | 78.83 (17.84) |
| Fbsp2-1-1 | MAE | 2.27 | 4.34 | 6.26 | 5.90 | 4.48 | 5.63 | 6.76 | 5.49 | 5.29 | 3.97 | 5.04 (1.31) |
| | MAPE (%) | 3.32 | 6.64 | 10.13 | 9.45 | 7.15 | 9.48 | 11.48 | 7.23 | 8.69 | 6.00 | 7.96 (2.37) |
| | RMSE | 3.07 | 5.22 | 7.04 | 6.75 | 5.32 | 6.48 | 7.51 | 6.38 | 6.17 | 4.85 | 5.88 (1.30) |
| | Prec (%) | 99.18 | 91.11 | 47.22 | 45.73 | 93.33 | 39.67 | 31.75 | 76.22 | 76.28 | 95.21 | 69.57 (25.91) |
| Shan1.5-1.0 | MAE | 3.36 | 4.46 | 5.76 | 5.61 | 3.60 | 4.81 | 5.50 | 6.57 | 4.97 | 4.18 | 4.88 (1.01) |
| | MAPE (%) | 4.90 | 6.78 | 9.74 | 9.56 | 5.65 | 8.60 | 9.89 | 9.03 | 8.14 | 6.21 | 7.85 (1.83) |
| | RMSE | 4.26 | 5.39 | 6.62 | 6.51 | 4.47 | 5.84 | 6.40 | 7.34 | 5.83 | 5.10 | 5.78 (0.98) |
| | Prec (%) | 95.36 | 83.37 | 63.97 | 70.44 | 94.54 | 58.51 | 70.68 | 41.26 | 82.33 | 89.82 | 75.03 (17.36) |

HR absolute errors for each wavelet method are represented as histograms in Figure 8.

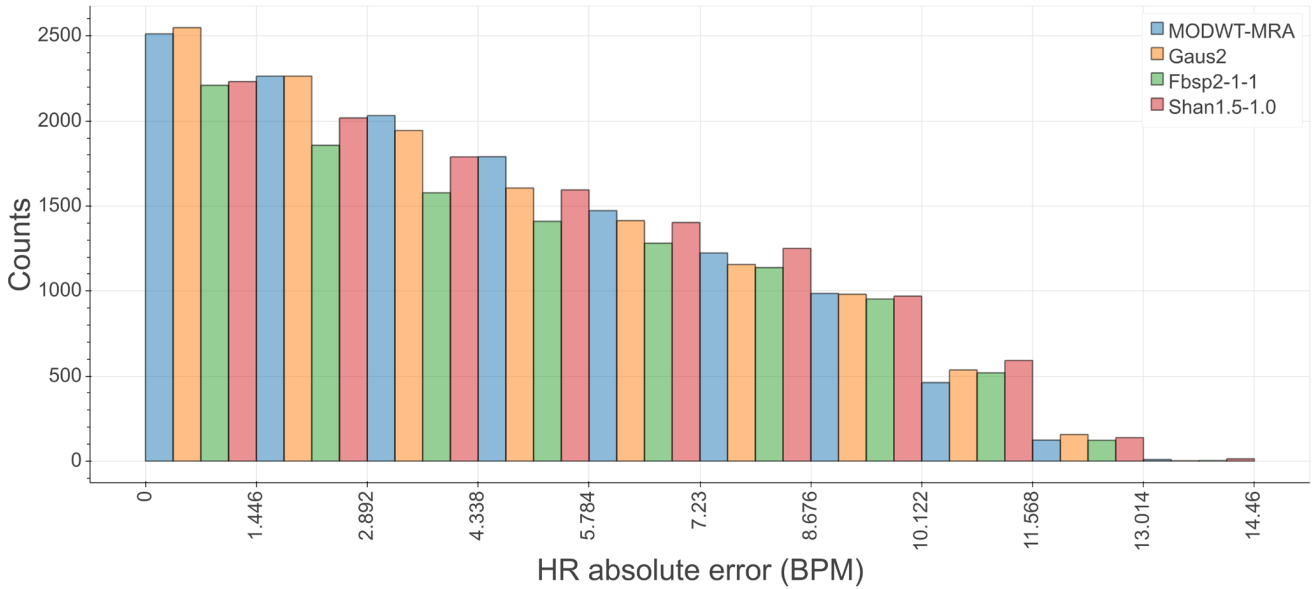

Figure 8 Distribution of the HR absolute error as histograms for each wavelet function across DataSet1.

It is clear from the figure that the HR detection performances of Gaus2 and MODWT-MRA were comparable. Moreover, Figure 9 shows the Bland-Altman plot of HR for Gaus2 function across DataSet1. The limits of agreement (LoA) were computed as described in [48,49] given the fact that multiple observations per individual are available. The upper and lower LoA values were 10.95 and -11.17 BPM ($r_{mr} = 0.38, P < .001$); $r_{mr}$ is the "repeated measures correlation" described in [50].

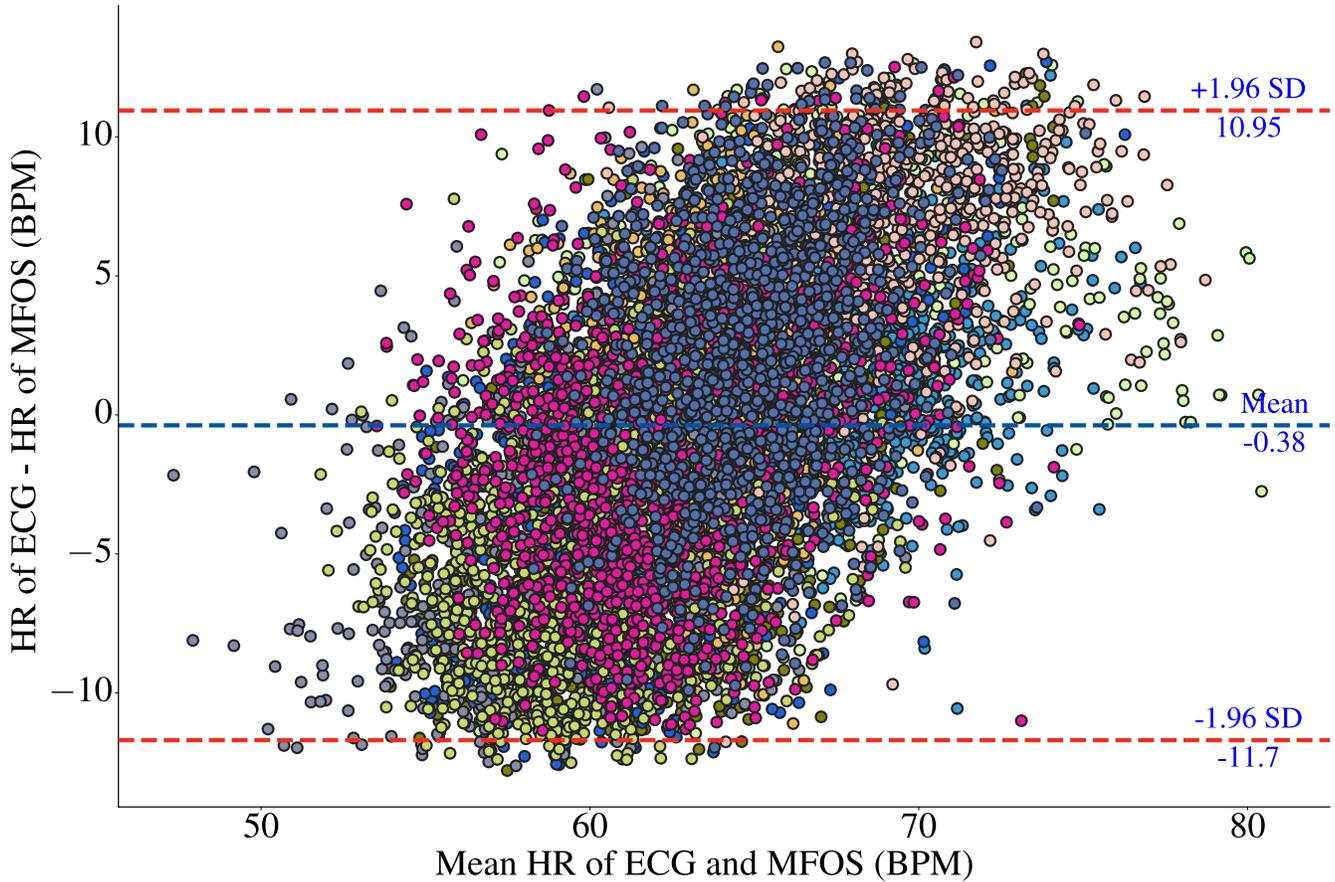

Figure 9 Bland-Altman plots of Gaus2 method across DataSet1. Markers' colors were randomly assigned for each patient.

### 4.2. Performance Evaluation of Heart Rate Detection: DataSet2

This particular dataset was challenging because BCG signals were gathered in a noisy environment. The signal quality was affected to a large degree by the massage chair's movement and loss of contact with the MFOS. Average error measures MAE, MAPE, and RMSE for **a)** MODWT-MRA, **b)** Gaus2, **c)** Fbsp-2-1-1, and **d)** Shan1.5-1.0 were: **a)** 4.95 (1.19), 7.57% (1.74%), and 5.77 (1.09); **b)** 4.85 (1.20), 7.45% (2.17%), and 5.66 (1.11); **c)** 5.19 (1.63), 8.04% (3.17%), and 5.96 (1.48); **d)** 4.95 (1.07), 7.57% (1.63%), and 5.80 (0.97), respectively. Overall, Gaus2 and Shan1.5-1.0 performed slightly better than MODW-MRA and Fbsp2-1-1. Similar to DataSet1, the largest error values were scored by Fpsp2-1-1, i.e., 9.58, 16.68%, and 9.73. Figure 10 illustrates the HR detection error of each wavelet-based function across all participants.

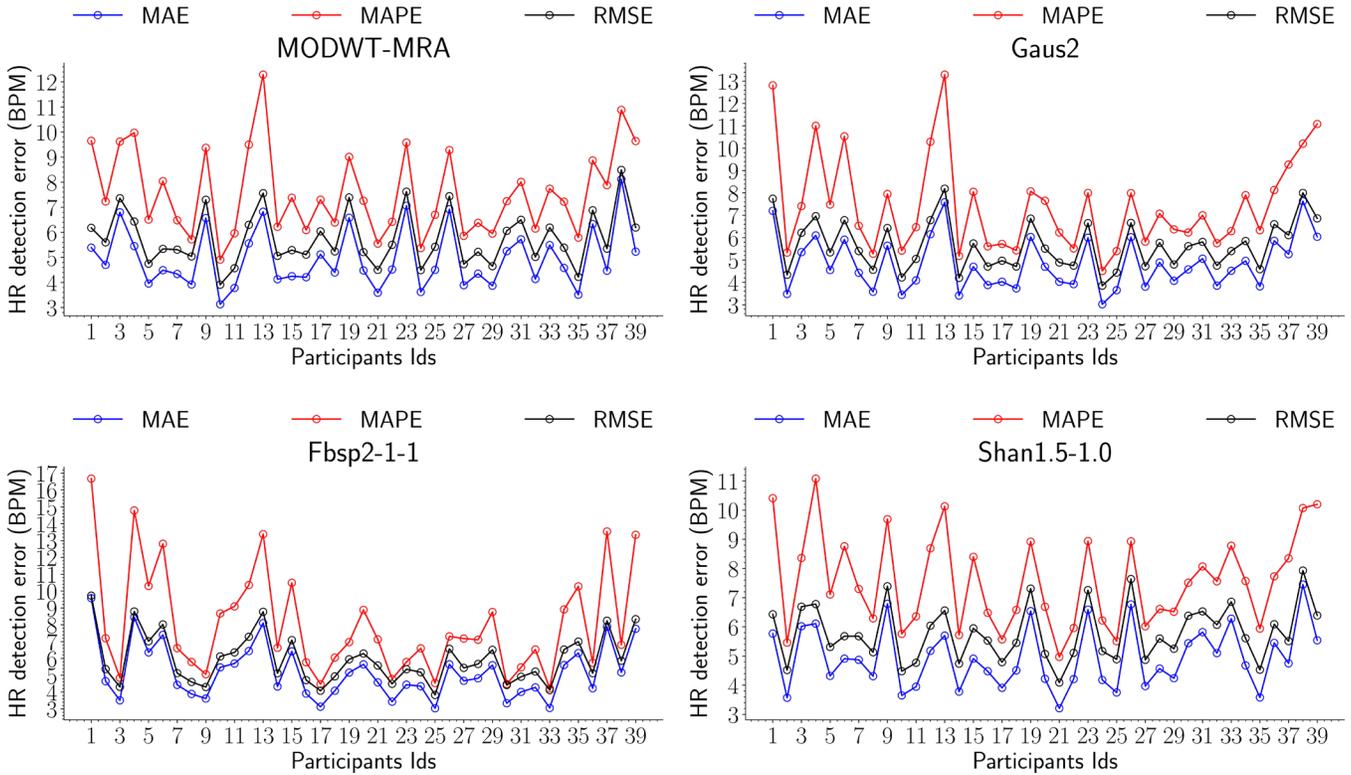

Figure 10 Overall performance measures (MAE, MAPE, RMSE) of the HR detection across DataSet2 using the 4 wavelet-based functions, i.e., MODWT-MRA, Gaus2, Fbsp2-1-1, and Shan1.5-1.0 (top left, top right, bottom left, and bottom right).

That said, Gaus2 scored the largest Prec value, i.e., 81.14% (14.36%), whereas the Prec values for MODWT-MRA, Fbsp2-1-1, and Shan1.5-1.0 were 77.12% (18.72%), 76.24% (23.68%), and 76.02% (14.63%), respectively. The maximum, minimum, and overall Prec values for each wavelet-based function are specified in Table 6.

Table 6 The maximum, minimum, and total values of the precision for the 4 wavelet-based functions (i.e., MODWT-MRA, Gaus2, Fbsp2-1-1, and Shan1.5-1.0) and the template matching approach.

| Methods | Precision (%) | | |
| --- | --- | --- | --- |
| | Minimum | Maximum | Mean (SD) |
| MODWT-MRA | 31.53 | 97.64 | 77.12 (18.72) |
| Gaus2 | 45.95 | 99.09 | 81.14 (14.36) |
| Fbsp2-1-1 | 21.14 | 99.12 | 76.24 (23.68) |
| Shan1.5-1.0 | 42.34 | 93.13 | 76.02 (14.63) |
| Template Matching | 26.24 | 92.86 | 72.83 (15) |

### 4.2.1. TM Related Results

The TM method's performance depends mainly on the template and similarity measure, i.e., a cross-correlation in our case. The BCG template was created from DataSet1. Accordingly, the intention was to utilize this template for HR detection across DataSet2. Overall, the TM achieved MAE, MAPE, and RMSE of 4.74 (0.66), 7.46% (1.42%), and 5.67 (0.68), respectively. As illustrated in Figure 11, the smallest values were 2.59, 4.04%, and 3.17 for participant 21, while the largest values were 5.96, 11.40%, and 6.84 for participant 13. Although obtained results were reasonable, the total Prec, i.e., 72.83% (15%), was not as good as Gaus2 (Table 6).

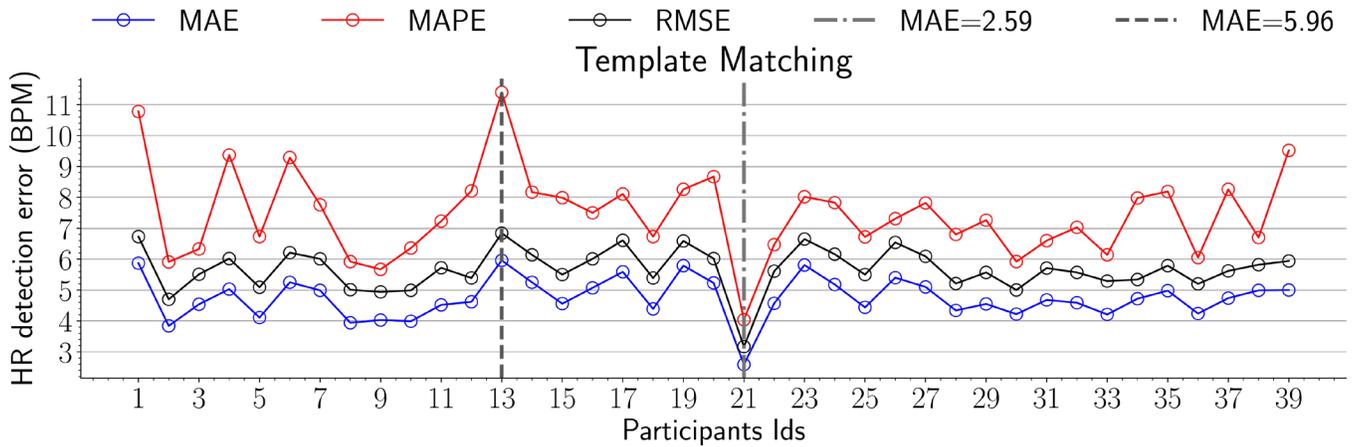

Figure 11 Overall performance measures (MAE, MAPE, RMSE) of the HR detection across DataSet2 using the template matching approach.

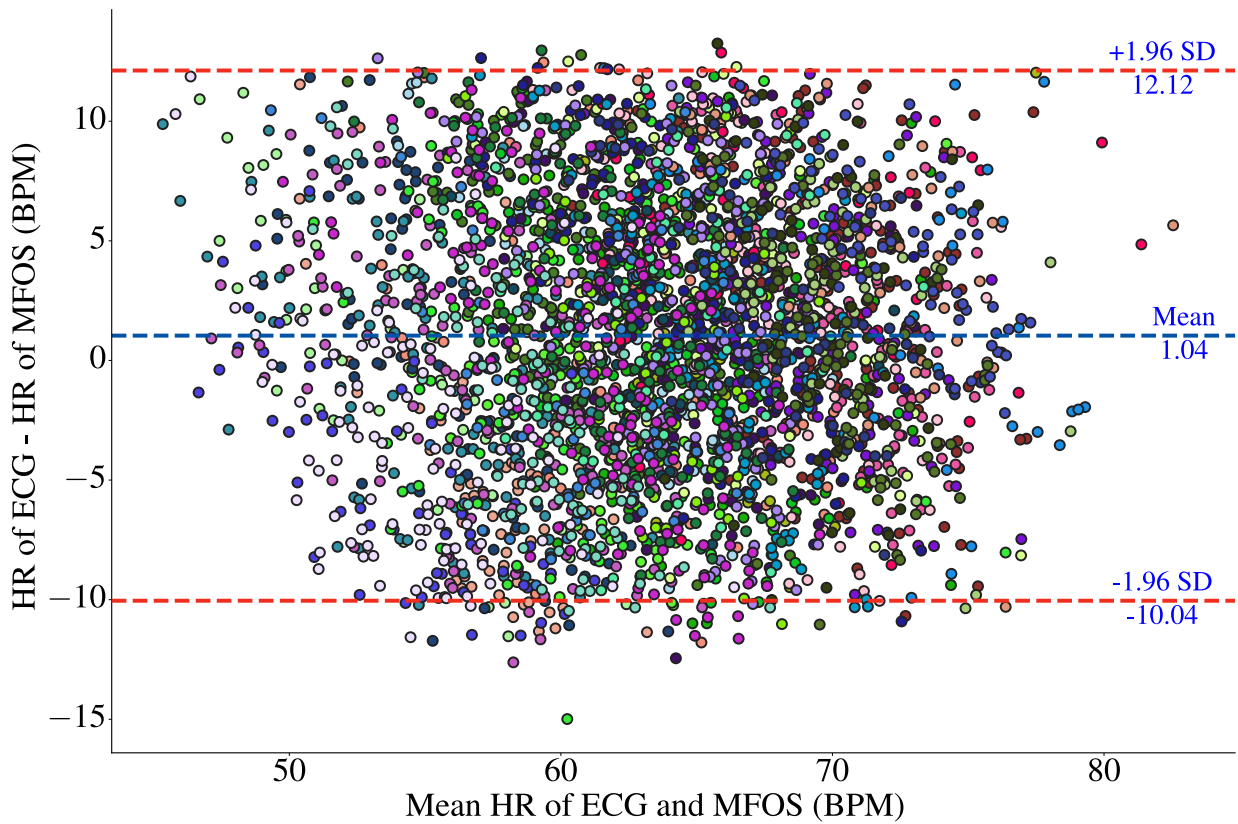

Figure 12 Bland-Altman plot of the TM approach across DataSet2. Markers' colors were randomly assigned for each subject.

Still, this fairly small Prec value was expected given the fact that the template was generated from a different dataset. Moreover, the BCG signals in DataSet2 were heavily corrupted by head movement artifacts. Figure 12 shows the Bland-Altman plot of HR for the TM approach across DataSet2. The upper and lower LoA values were 12.12 and -10.04 BPM ($r_{mr} = 0.34, P < .001$). Similarly, it could be seen from the plot the relatively small number of HR points (i.e., Prec) in contrast to Gaus2. Additionally, $r_{mr}$ was 0.38; however, for Gaus2 it was 0.37. Figure 13 shows the HR distribution for Gau2 and TM versus the reference ECG for the second participant across DataSet2. In addition, Figure 14 shows the boxplots with p-values for Gaus2 and TM versus the reference ECG across DataSet2.

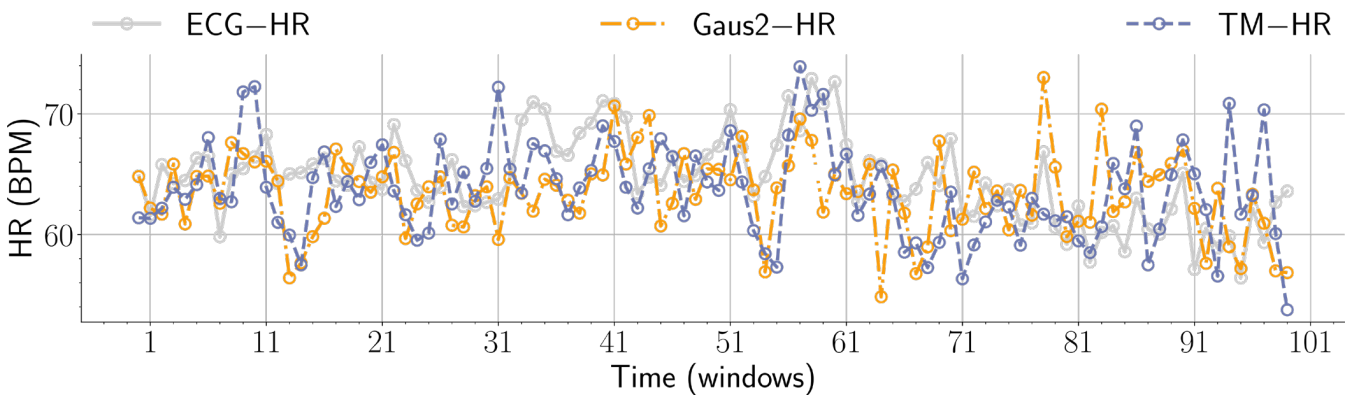

Figure 13 HR distribution for the reference ECG, Gau2, and TM methods for participant 2 (DataSet2). Time-windows were included in the diagram if they had been evaluated by both methods.

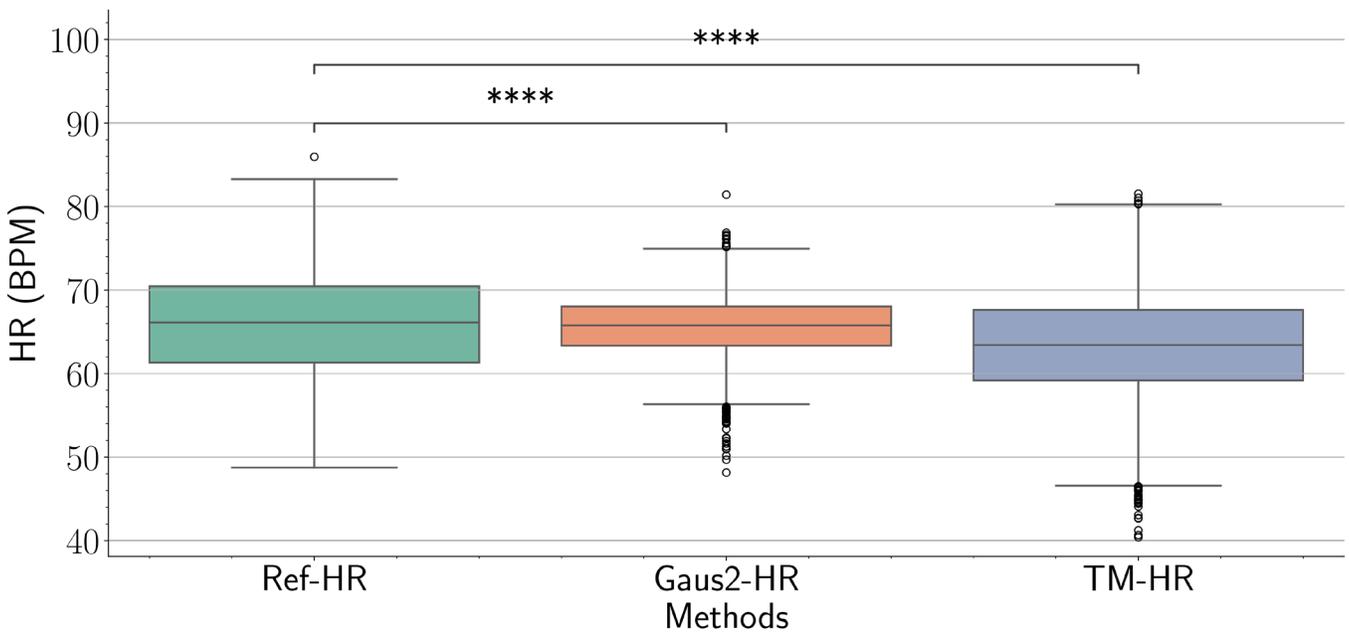

Figure 14 Boxplots with p-values for Gaus2 and TM methods vs. the reference ECG across DataSet2.

## 4.3. Performance Evaluation of Heart Rate Detection: DataSet3

For further checking the effectiveness of the TM approach, we used the BCG template generated from DataSet1 to detect HR in DataSet3. We down-sampled the FBG signals to 50Hz so that cardiac cycles could match the BCG template. As given in Table 7, reasonable results were obtained, in which the total MAE, MAPE, and RMSE were 3.43 (1.26), 5.51% (2.31%), and 4.58 (1.26), respectively.

Table 7 HR detection performance metrics for DataSet3 using the TM approach.

| Metrics | Subjects Ids | | | | | | | | | | Mean (SD) |
|---|---|---|---|---|---|---|---|---|---|---|---|
| | 1 | 2 | 3 | 4 | 5 | 6 | 7 | 8 | 9 | 10 | |
| MAE | 3.18 | 2.76 | 5.62 | 4.43 | 3.02 | 2.89 | 1.42 | 4.79 | 2.25 | 3.93 | 3.43 (1.26) |
| MAPE (%) | 4.40 | 4.63 | 10.17 | 7.26 | 5.27 | 4.82 | 1.83 | 6.57 | 3.38 | 6.80 | 5.51 (2.31) |
| RMSE | 4.33 | 3.88 | 6.34 | 5.24 | 5.71 | 4.19 | 2.37 | 5.51 | 2.99 | 5.22 | 4.58 (1.26) |
| Prec (%) | 92.55 | 60.44 | 54.44 | 87.78 | 81.4 | 88.04 | 96.7 | 85.87 | 93.48 | 68.09 | 80.88 (14.72) |

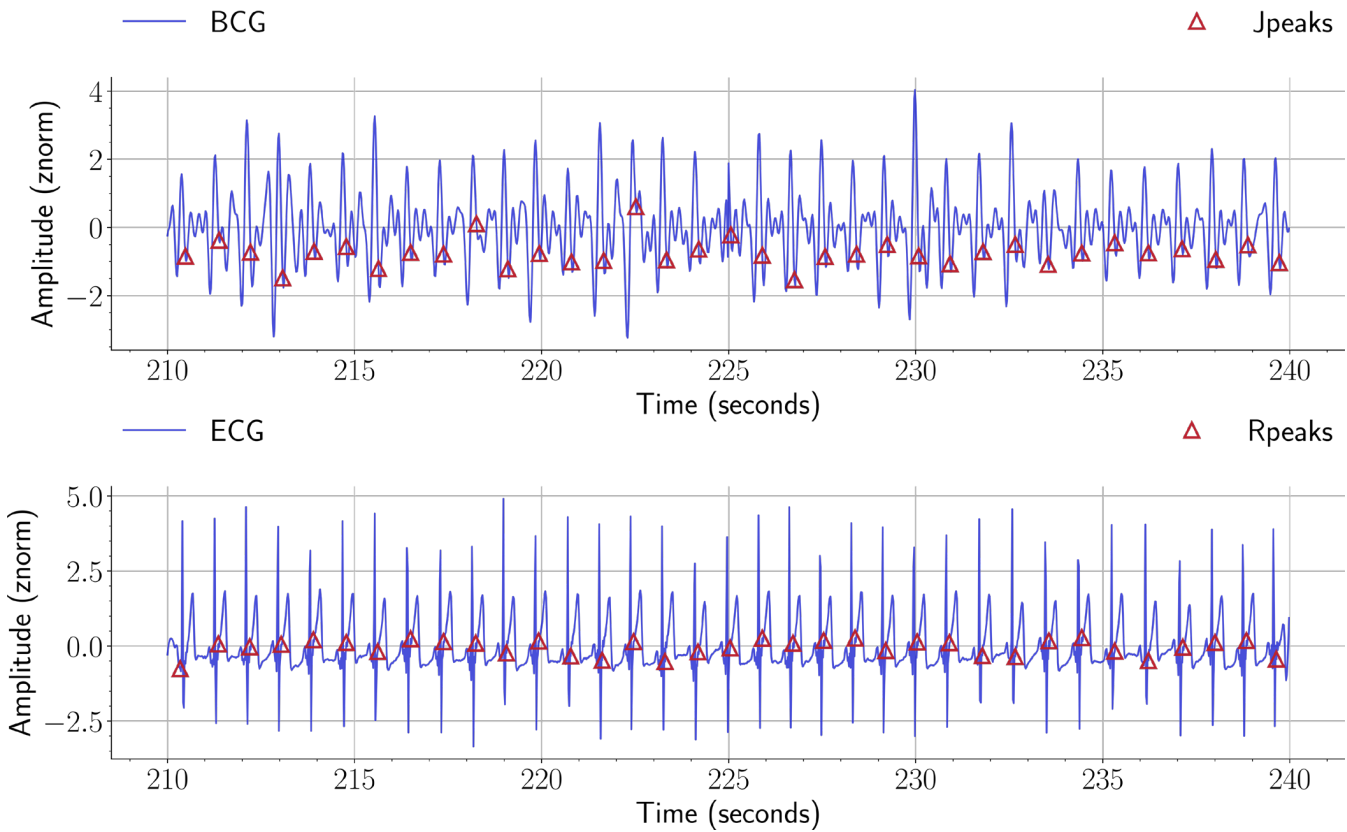

Figure 15 HR detection results using the TM approach (subject 1, DataSet3). The top figure shows a 30-second BCG signal with the J-peaks annotated by up-pointing triangles. The bottom figure shows the corresponding ECG signal with the R-peaks annotated by up-pointing triangles.

Figure 15 demonstrates the performance of the TM approach for J-peaks detection across DataSet3. The top section of the figure shows a 30-second BCG signal with the J-peaks marked by up-pointing triangles. Besides, the bottom section of the figures shows the equivalent ECG signal with the R-peaks labeled by up-pointing triangles. Figure 16 shows the repeated measures correlation (Rmcorr) plot for HR detection across DataSet3 using the TM method. Across the ten subjects, Rmcorr and p-value were: $r_{mr} = 0.39 \ and \ P < .001$.

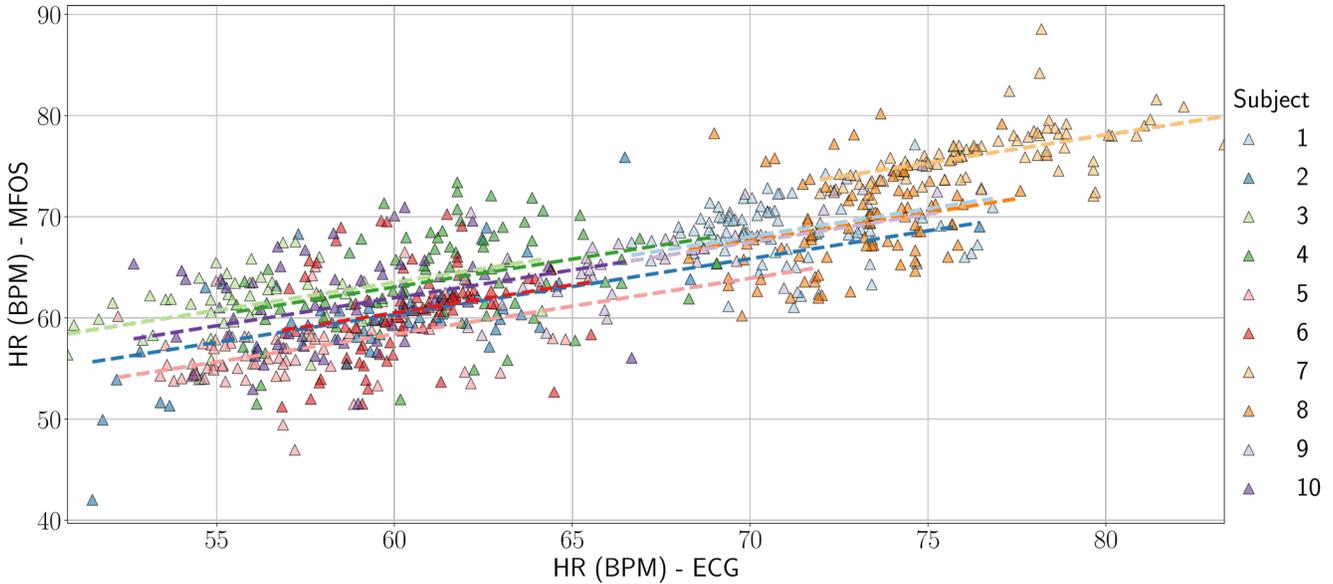

Figure 16 Repeated measures correlation (rmcorr) coefficient plot [50] for HR detection across DataSet3 using the TM method.

### 4.4. Performance Evaluation of Heart Rate Detection: DataSet4

Firstly, BCG signals obtained from EMFis and load cells were down-sampled to 50Hz. For the EMFis, the overall MAE, MAPE, RMSE were 2.15 (2.33), 2.91% (2.82%), and 2.62 (2.42) using Gaus2. For the TM, the total MAE, MAPE, RMSE were 4.07 (2.35), 5.79% (2.91%), and 4.76 (2.29), respectively. For the load cells, the total MAE, MAPE, RMSE were 1.63 (1.40), 2.28% (1.78%), and 2.14 (1.62) for Gaus2. Besides, the total MAE, MAPE, RMSE were 3.98 (2.18), 6.12% (3.77%), and 4.67 (2.20).

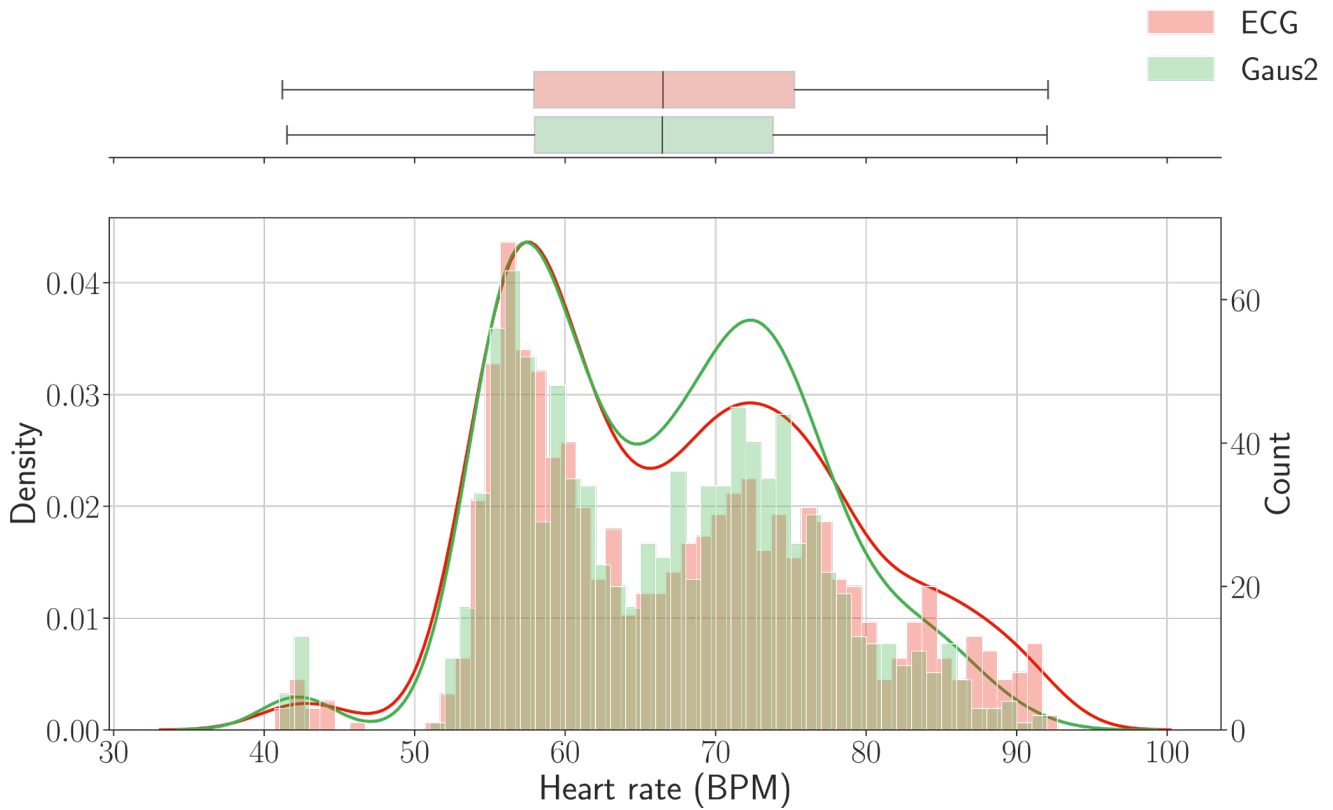

*Figure 17 Superimposed histograms of the ECG-derived HR and Gaus2-derived HR across DataSet4 (EMFis). HR values are represented as boxplots on top of the figure.*

Figure 17 shows superimposed histograms of the ECG-derived HR and Gaus2-derived HR across DataSet4 for the EMFis. Furthermore, the HR-derived values for each method are represented as boxplots on top of the figure to show the distribution of the data. On the other hand, the overall performance measures of the HR detection across DataSet4 using Gaus2 are shown in the top left and top right of Figure 18. Similarly, the performance measures across DataSet4 using TM are shown in the bottom left and bottom right of Figure 18.

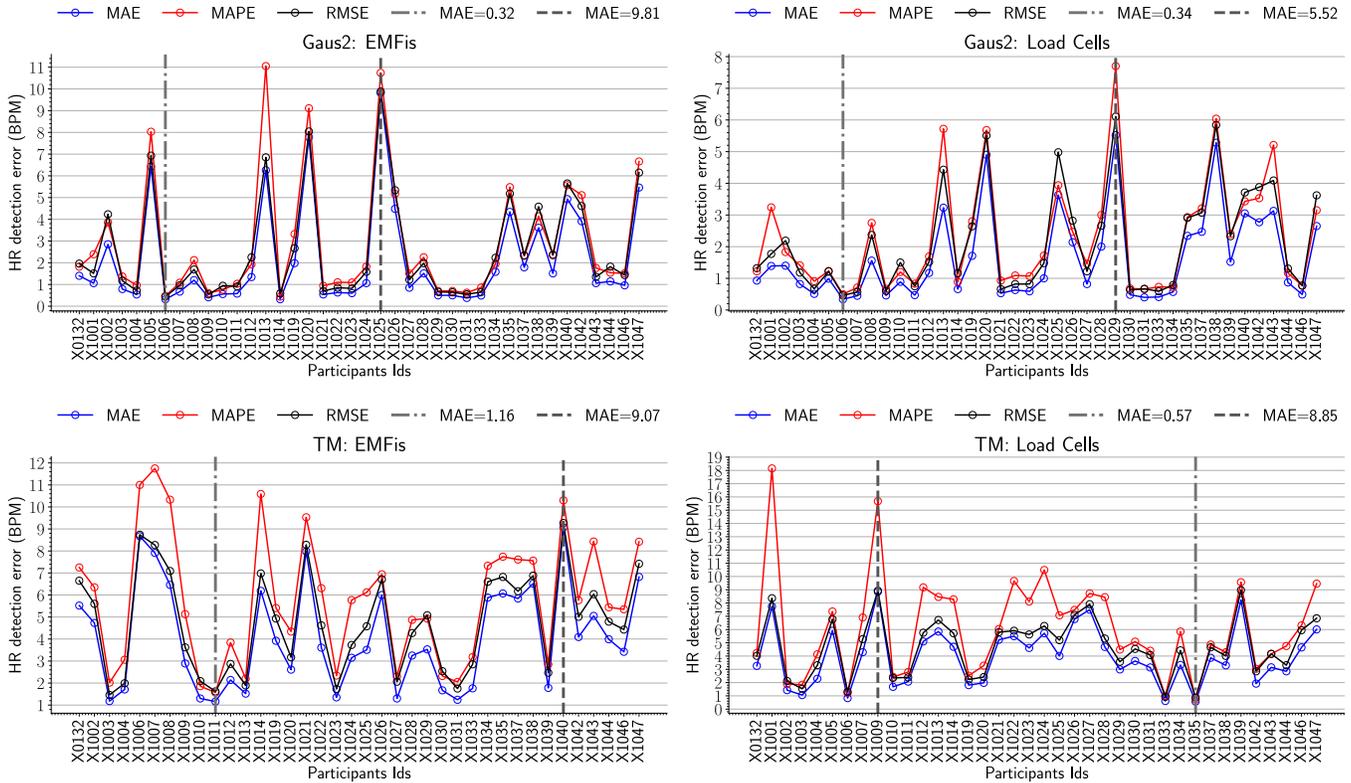

Figure 18 Overall performance measures (MAE, MAPE, RMSE) of the HR detection across DataSet4 using Gaus2 (top left and top right) and TM (bottom left and bottom right) methods.

## 5. Discussion

### 5.1. DataSet1

Gaus2 seemed to provide information about cardiac cycles more accurately than other wavelets. The total Prec accomplished by Gaus2, i.e., 78.83% (17.84%), was slightly inferior to MODWT-MRA, i.e., 80.22% (19.01%). Nonetheless, Gaus2 generated more favorable results with respect to the error measures (Table 5). These results make sense because the scales are discretized more finely than the MODWT. That said, MODWT-MRA requires less time for analyzing a 30-second BCG signal than Gaus2 (Table 3), which makes it a suitable candidate for real-time analysis.

### 5.2. DataSet2

It is important to highlight that our findings for the Gaus2 and TM methods provided accepted overall aggregated results that were less than 10% MAPE. As stated in [51], an error rate of $\pm 10\%$ can be regarded as an accurate threshold for medical ECG monitors. To this end, preferring one method over the other will depend on the application requirements. To illustrate, TM method produced a little better result in detecting HR than Gaus2 (Figure 13 and Figure 14). Yet, the total Prec was smaller than Gaus2. Thus, the TM approach can be more practical for HR detection in a well-controlled environment in which motion artifacts can be minimized. On the other hand, the Gaus2 method seems less susceptible to motion artifacts. Hence, it can be more practical in real-life situations. Still, selecting an optimal wavelet function and scale requires prior knowledge about the BCG signal morphology. These two parameters will vary from one specific sensor to another. Besides, it should be pointed out that the HR detection results can differ significantly from one scale to another. That is to say, a particular scale or wavelet can only provide adequate results for individual cases while the opposite happens for other cases. A situation like this occurred, for example, with Fbsp2-1-1 in DataSet1 (Section 4.1).

### 5.3. DataSet3

As given by Table 7, the total Prec was acceptable, i.e., 80.88% (14.72%). This Prec value was fairly similar to a previous work in which the fusion was performed in the frequency domain using the cepstrum, and the total Prec reported was 84% [52]. In summary, these results may support the potential of using a BCG template from a particular dataset to detect HR in a different dataset and under different conditions. The three methods described thus far have provided consistent results for HR detection from BCG signals. Moreover, the total Prec values achieved by the three methods were fairly reasonable, considering that BCG signals were recorded in non-restricted environments; in other words, subjects' movements were allowed.

### 5.4. DataSet4

For the EMFis, Gaus2 achieved closer results to the reference ECG than the TM (Figure 17). Moreover, the total Prec attained by Gaus2, i.e., 93.65% (16.57%), was better than that of the TM, i.e., 84.08% (20.89%). Rmcorr and p-value were: $r_{mr} = 0.65, P < .001$ and $r_{mr} = 0.35, P < .001$ for Gaus2 and TM, respectively. On the other hand, the TM failed to analyze signals from subjects "X1001" and "X1005" because the signal's morphology was quite different compared to the MFOS. For the load cells, Gaus2 outperformed the TM in a similar way as it had done using the EMFis. Unlike Gaus2, the TM failed to analyze signals from two subjects, i.e., "X1008", "X1040". Rmcorr and p-values were: $r_{mr} = 0.66, P < .001$ and $r_{mr} = 0.36, P < .001$ for Gaus2 and TM, accordingly. The results mentioned above (Figure 18) demonstrate the superiority of Gaus2 for HR detection compared with the TM. Table 8 (Appendix) shows the Rmcorr and LoA achieved by the three methods across the four datasets.

## 6. Conclusion

The primary objective of this comparative study was to investigate the performance of three HR detection algorithms (MODWT-MRA, CWT, and TM) across several datasets and under different experimental setups. To this goal, we tested the three algorithms using two local datasets and one publicly available dataset. Besides, the performance of each method was assessed using four error measures, namely MAE, MAPE, RMSE, and Prec. For MODWT-MRA, wavelet bior3.9 with four decomposition levels were adopted. Besides, the 4[th] level smooth coefficient was assigned for detecting cardiac cycles. For CWT, three wavelets were analyzed, i.e., Gaus2 (20[th] scale of 30), Fbsp2-1-1 (45[th] scale of 100), and Shan1.5-1.0 (75[th] scale of 100). In general, CWT-Gaus2 achieved more favorable outcomes compared with other wavelets. Regarding the TM method, a BCG template was generated from DataSet1 as a training set. Then, it was used for HR detection in the remaining datasets. Satisfactory results were achieved for DataSet1 and DataSet2. Nevertheless, it was unsuccessful to analyze two subjects out of 40 from DataSet4 because the template was created from a completely independent dataset. These results show the shortcoming of the TM method for analyzing BCG signals collected by different sensors. On the other hand, wavelet-based methods have proved successful regardless of sensor types or experimental setups. The proposed methods were implemented on a Raspberry Pi to test their effectiveness for real-time applications. As a result, the average time required to analyze a 30-second BCG signal was less than one second for all methods. However, the MODWT-MRA had the highest performance, with an average time of 0.04 seconds.

## Declaration of Competing Interest

None.

# Appendix

Table 8 Rmcorr, upper LoA, and lower LoA for wavelet-based methods and template matching across the four datasets.

| Methods | Rmcorr | Upper LoA | Lower LoA |
|---|---|---|---|
| | DataSet1 | | |
| MODWT-MRA | 0.38 | 12.13 | -10.30 |
| Gaus2 | 0.38 | 10.95 | -11.70 |
| Fbsp2-1-1 | 0.39 | 8.41 | -13.27 |
| Shan1.5-1.0 | 0.32 | 11.03 | -12.06 |
| Methods | DataSet2 | | |
| MODWT-MRA | 0.35 | 12.93 | -9.31 |
| Gaus2 | 0.31 | 11.61 | -11.16 |
| Fbsp2-1-1 | 0.25 | 8.82 | -13.61 |
| Shan1.5-1.0 | 0.27 | 12.09 | -10.80 |
| TM | 0.34 | 12.12 | -10.04 |
| Methods | DataSet3 | | |
| TM | 0.39 | 9.19 | -9.70 |
| Methods | DataSet4 | | |
| Gau2 (EMFis) | 0.65 | 7.76 | -5.39 |
| TM (EMFis) | 0.35 | 11.59 | -6.30 |
| Gaus2 (Load cells) | 0.66 | 5.06 | -4.88 |
| TM (Load cells) | 0.36 | 10.22 | -10.08 |

Table 9 All the errors (i.e., MAE, MAPE, RMSE, and Prec) reported for each of the datasets (i.e., DataSet1, DataSet2, DataSet3, and DataSet4) and each of the algorithms used (i.e., MODWT-MRA, CWT-Gaus2, CWT-Fbsp2-1-1, CWT-Shan1.5-1.0, TM).

|  | MODWT-MRA | CWT-Gaus2 | CWT-Fbsp2-1-1 | CWT-Shan1.5-1.0 | TM |
|---|---|---|---|---|---|
|  | | | DataSet1 | | |
| MAE | 4.71 (1.07) | 4.71 (1.22) | 5.04 (1.31) | 4.88 (1.01) | |
| MAPE (%) | 7.61 (1.65) | 7.58 (2.17) | 7.96 (2.37) | 7.85 (1.83) | |
| RMSE | 5.59 (1.02) | 5.58 (1.20) | 5.88 (1.30) | 5.78 (0.98) | |
| Prec (%) | 80.22 (19.01) | 78.83 (17.84) | 69.57 (25.91) | 75.03 (17.36) | |
|  | | | DataSet2 | | |
| MAE | 4.95 (1.19) | 4.85 (1.20) | 5.19 (1.63) | 4.95 (1.07) | 4.74 (0.66) |
| MAPE (%) | 7.57 (1.74) | 7.45 (2.17) | 8.04 (3.17) | 7.57 (1.63) | 7.46 (1.42) |
| RMSE | 5.77 (1.09) | 5.66 (1.11) | 5.96 (1.48) | 5.80 (0.97) | 5.67 (0.68) |
| Prec (%) | 77.12 (18.72) | 81.14 (14.36) | 76.24 (23.68) | 76.02 (14.63) | 72.83 (15) |
|  | | | DataSet3 | | |
| MAE | | | | | 3.43 (1.26) |
| MAPE | | | | | 5.51 (2.31) |
| RMSE | | | | | 4.58 (1.26) |
| Prec (%) | | | | | 80.88 (14.72) |
|  | | | DataSet4 (EMFis) | | |
| MAE | | 2.15 (2.33) | | | 4.07 (2.35) |
| MAPE (%) | | 2.91 (2.82) | | | 5.79 (2.91) |
| RMSE | | 2.62 (2.42) | | | 4.76 (2.29) |
| Prec (%) | | 93.65 (16.57) | | | 84.08 (20.89) |
|  | | | DataSet4 (Load Cells) | | |
| MAE | | 1.63 (1.40) | | | 3.98 (2.18) |
| MAPE (%) | | 2.28 (1.78) | | | 6.12 (3.77) |
| RMSE | | 2.14 (1.62) | | | 4.67 (2.20) |
| Prec (%) | | 95.56 (12.51) | | | 87.06 (19.37) |

## Technical Implementation of the MFOS

The MFOS is integrated into ambient assisted living (AAL) platforms known as UbiSmart [53,54] and AmI-IoT [55]. The working principle of the MFOS and its integration into the platforms can be found in [7,53–56]. In summary, the MFOS is considered as another sensor that contributes to the knowledge base of the AAL platform (e.g., UbiSmart). The MFOS communicator unit is wired to our Gateway (Figure 19).

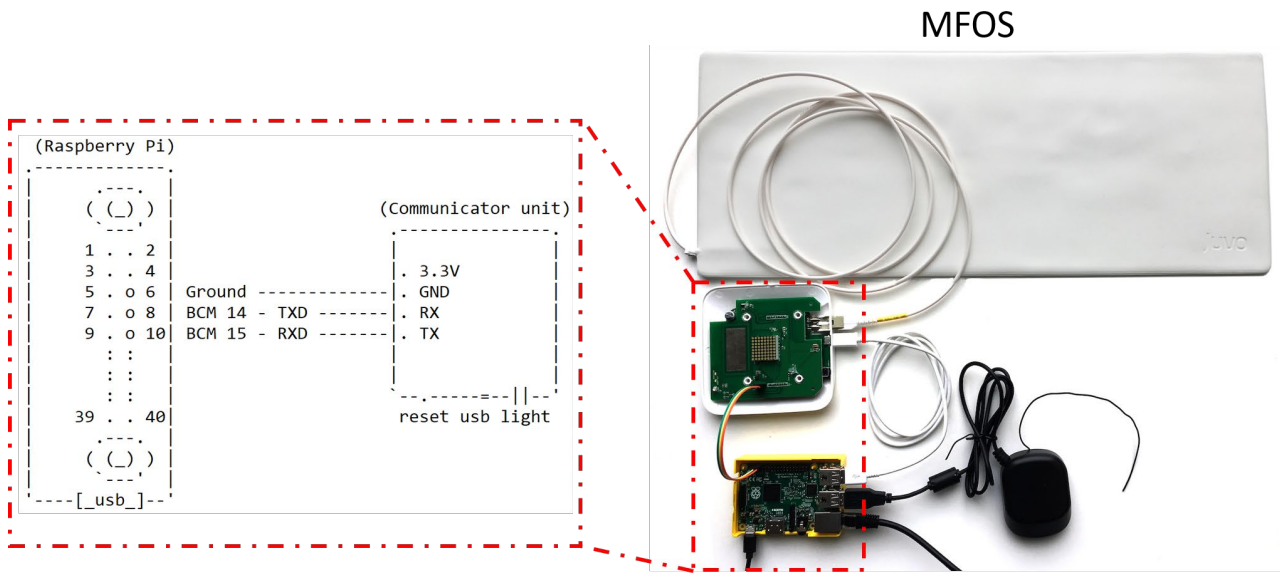

Figure 19 The connection between the MFOS and the RPi, i.e., the Gateway.

Voluminous raw data is read and stored on a micro-SD card for a deeper off-line analysis. Simultaneously, the data is preprocessed to generate high-level events, such as bed empty, bed motion, sleep. Currently, it operates on a time window of 10 seconds. For each time window, an event is produced. The events are then sent to the Server as structured sensor data using MQTT protocol over an Internet connection. The Server handles the received structured information (event). The MFOS will appear in the home description interface as available for the association to a house. If confirmed, this association is stored in the knowledge base (KB). Any subsequent events are then inserted into the KB of the associated house, allowing the reasoning engine to be aware of bed occupancy with respect to our ontology (Figure 20). In parallel, the raw data is processed every 5 minutes to extract information about the subject's respiratory effort and heart rate. This information is also inserted into the KB. Figure 20 shows how we can get raw sensor data from the MSOF via a RPi.

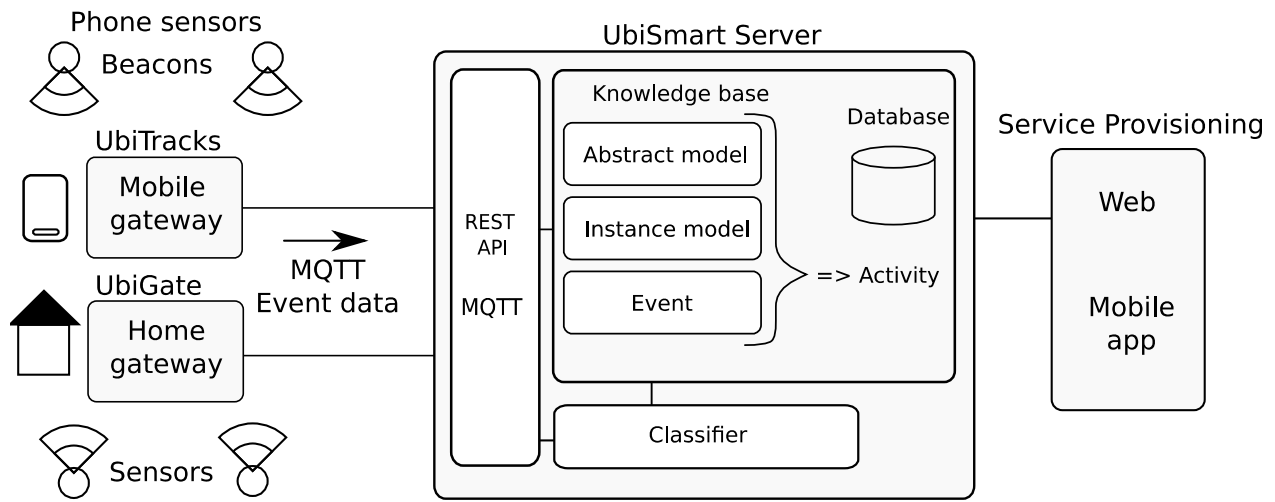

Figure 20 UbiSmart framework in three sections. The UbiGate, the Server, and the Service Provisioning [57].

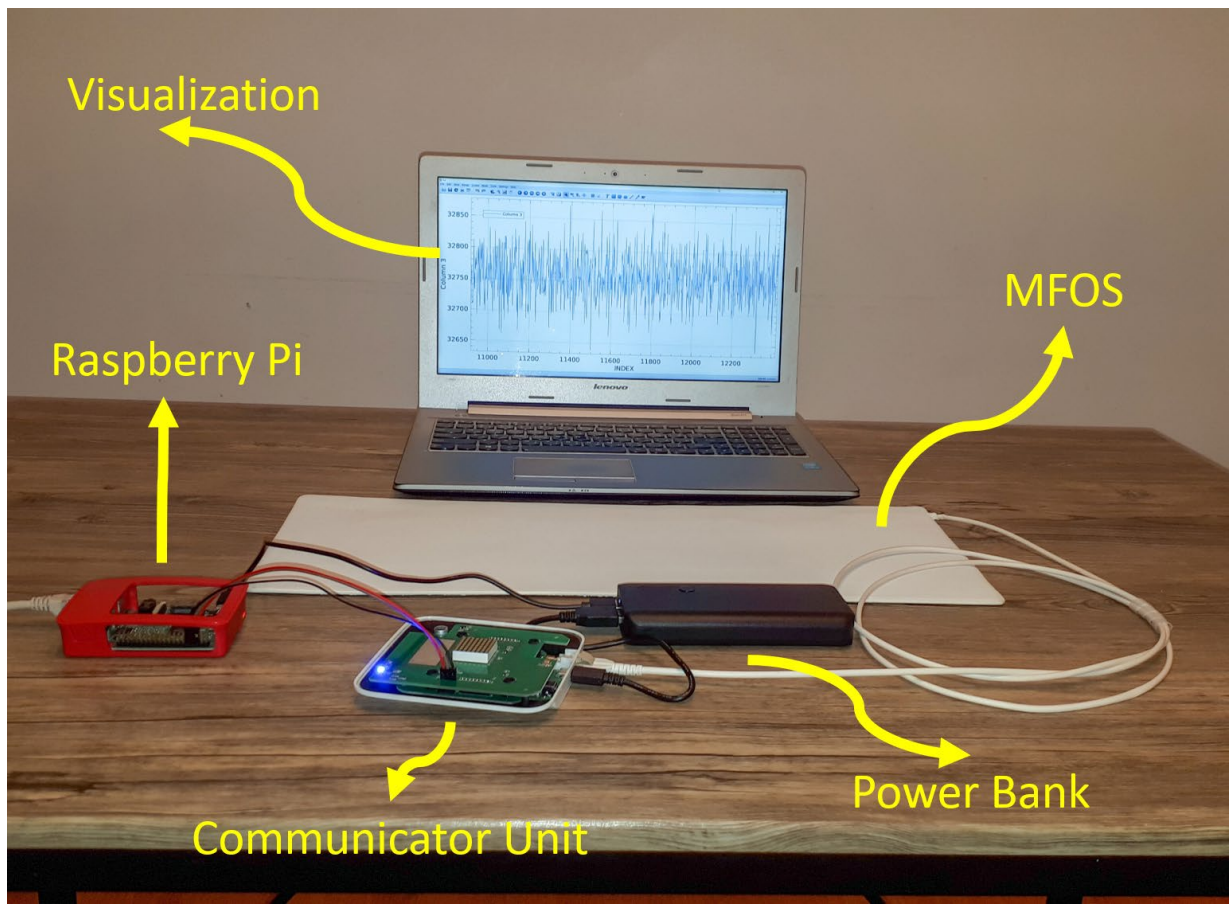

Figure 21 An experimental setup for acquiring raw sensor data from the MFOS. The data were obtained by connecting the communicator unit to the GPIO (general-purpose input/output) of the RPi. Then, the Pi is connected to a PC through an SSH connection. The KST[3] plotting tool (implemented on the RPi) was used to display the data in real-time.

---

[3] https://kst-plot.kde.org/